\documentclass[aps,pre, twocolumn, superscriptaddress, balancelastpage, showpacs, reprint,nofootinbib]{revtex4-2}

\usepackage[T1]{fontenc}
\usepackage{blindtext}
\usepackage{centernot}
\usepackage{graphicx}
\usepackage{amsmath,bbold}
\usepackage{times}
\usepackage{amssymb}
\usepackage{mathrsfs}
\usepackage{chemarr}
\usepackage{color}
\usepackage{url}
\usepackage{version}
\usepackage[hidelinks]{hyperref}
\usepackage{mwe,tikz}
\usepackage[percent]{overpic}
\usepackage{bm}
\usepackage[export]{adjustbox}
\definecolor{linkcolor}{rgb}{0,0,0.6} 
\definecolor{forestgreen}{rgb}{0.13, 0.55, 0.13}
\definecolor{frenchblue}{rgb}{0.0, 0.45, 0.73}
\definecolor{burntsienna}{rgb}{0.91, 0.45, 0.32}
\usepackage{algorithm}
\usepackage[noend]{algpseudocode}
\usepackage{array}
\usepackage{multirow}

\usepackage{enumitem}
\usepackage{cases}
\usepackage[normalem]{ulem}
\usepackage{transparent}

\newcommand\rsout{\bgroup\markoverwith{\textcolor{red}{\rule[0.5ex]{2pt}{0.8pt}}}\ULon}

\newcommand{\rhostar}{{\rho^\star}}

\newcommand{\dd}{\mathrm{d}}
\newcommand{\ee}{\mathrm{e}}

\newcommand{\alphap}{{\alpha_\mathrm{p}}}

\newcommand{\bmr}{\bm r}

\newcommand{\rational}{\Gamma}

\DeclareMathOperator*{\sgn}{sgn}
\DeclareMathOperator*{\tr}{tr}
\DeclareMathOperator\arctanh{arctanh}

\usepackage{lipsum}
\usetikzlibrary{patterns}

\newcolumntype{P}[1]{>{\centering\arraybackslash}p{#1}}
\newcolumntype{M}[1]{>{\centering\arraybackslash}m{#1}}

\begin{document}
  

\title{Socioeconomic agents as active matter in  nonequilibrium Sakoda-Schelling models}

\author{Ruben Zakine}
\email{ruben.zakine@polytechnique.edu}
\affiliation{Chair of Econophysics and Complex Systems, \'Ecole polytechnique, 91128 Palaiseau Cedex, France}
\affiliation{LadHyX, CNRS, École polytechnique, Institut Polytechnique de Paris, 91120 Palaiseau, France}

\author{Jérôme Garnier-Brun}
\affiliation{Chair of Econophysics and Complex Systems, \'Ecole polytechnique, 91128 Palaiseau Cedex, France}
\affiliation{LadHyX, CNRS, École polytechnique, Institut Polytechnique de Paris, 91120 Palaiseau, France}

\author{Antoine-Cyrus Becharat}
\affiliation{Chair of Econophysics and Complex Systems, \'Ecole polytechnique, 91128 Palaiseau Cedex, France}
\affiliation{LadHyX, CNRS, École polytechnique, Institut Polytechnique de Paris, 91120 Palaiseau, France}

\author{Michael Benzaquen}
\affiliation{Chair of Econophysics and Complex Systems, \'Ecole polytechnique, 91128 Palaiseau Cedex, France}
\affiliation{LadHyX, CNRS, École polytechnique, Institut Polytechnique de Paris, 91120 Palaiseau, France}
\affiliation{Capital Fund Management, 23 Rue de l’Universit\'e, 75007 Paris, France}

\date{\today}

\begin{abstract}
How robust are socioeconomic agent-based models with respect to the details of the agents' decision rule? 
We tackle this question by considering an occupation model in the spirit of the Sakoda-Schelling model, historically introduced to shed light on segregation dynamics among human groups. For a large class of utility functions and decision rules, we pinpoint the nonequilibrium nature of the agent dynamics, while recovering the equilibrium-like phase separation phenomenology. 
Within the mean-field approximation we show how the model can be mapped, to some extent, onto an active matter field description. 
Finally, we consider non-reciprocal interactions between two populations, and show how they can lead to non-steady macroscopic behavior. We believe our approach provides a unifying framework to further study geography-dependent agent-based models, notably paving the way for joint consideration of population and price dynamics within a field theoretic approach. 
\end{abstract}

\maketitle 

\section{Introduction}

Will a collective system in which individuals share the same common goal ever reach an optimal state? This nontrivial question is at the very core of strong debates among economists, notably because the notion of ``optimal state'' is intrinsically political and most often ill-defined. 
Despite the common idea that a system made of agents individualistically improving their outcome will spontaneously converge by the action of the ``invisible hand'' to an optimal collective state, simple models have been shown to contradict this belief~\cite{hotelling1929stability,pigou2017economics,braess1968paradoxon}.
A well documented example of such system is the celebrated Schelling model~\cite{schelling1971}. The latter can be considered to be a variant of the model previously\footnote{To be perfectly precise, the first mention of Sakoda's model can be traced back to his unpublished PhD thesis completed in 1946, while Schelling's work can be found in a 1969 working paper \cite{hegselmann_thomas_2017}. In any case, there is no reason to think either author took inspiration from the other, the objective of the papers being clearly quite different.}  introduced by Sakoda \cite{sakoda_checkerboard_1971}, and will thus be referred henceforth as the Sakoda-Schelling model. To understand some aspects of urban segregation in post-WWII American cities, and more widely of urban and social dynamics, both authors proposed simple lattice models of idealized cities. Each site, representing an accommodation, can be empty or occupied by an agent belonging to one of two sub-populations in the system. Interestingly, Schelling observed that when introducing a slight preference for agents to be surrounded by neighbors of their own group, the system evolves towards configurations with completely segregated regions. While in fact not very well suited to explain urban segregation, which is intimately related to past and present public policies rather than self-organization \cite{boustan2013racial,trounstine2018segregation}, the model illustrates how the \textit{micromotives} of the agents may lead to unanticipated \textit{macrobehavior}~\cite{schelling2006micromotives}. 

Along the years, the Sakoda-Schelling model has attracted further attention of statistical physicists~\cite{vinkovic_physical_2006, dallasta_marsili_2008, gauvin2009_phaseDiag, rogers2011unified}, due to its simple microscopic rules, its paradoxical macroscopic consequences and its unconventional non-local particle moves. To the usual end of bridging the gap from \emph{micro to macro}, mappings onto equilibrium systems were suggested~\cite{gauvin_BEG_2010}, but with limited analytical results.
To gain a more in-depth understanding of the mechanism through which individual choices may lead to sub-optimal collective outcomes, Grauwin \textit{et al.} introduced a modified version of the Schelling model with a single type of agent occupying a lattice divided in pre-defined neighborhoods, or blocks \cite{grauwin_bertin_2009}. In this occupation model, the agents now base their decisions on the neighborhood density, which is identical for all the agents in a given block. This fixed neighborhood structure then allows to describe analytically the steady state as the minimizer of a free energy, and to recover a nontrivial phase with suboptimal jam-packed neighborhoods.
Subsequent works have then explored variations of these different models focusing on the effect of altruistic agents \cite{jensen_catalytic_2018}, dynamics close to criticality~\cite{barmpalias2015tipping,ortega2021schelling,ortega2021avalanches} or habit formation \cite{abella2022aging}.

Even in the seemingly simpler occupation problem of Grauwin \textit{et al.}~\cite{grauwin_bertin_2009}, several questions persist,  both from the socioeconomic and statistical physics perspectives. In particular, the role of the specific decision rule and the precise nature of neighborhoods on the phenomenology of the model remain unclear. Indeed, to allow for the standard techniques of statistical mechanics to be applicable, the choice of the neighborhoods and the dynamics is very constrained, see \cite{bouchaud_crises_2013}. As will be discussed in detail, most non-trivial decision rules lead the system out of thermodynamic equilibrium, requiring calculations that are not always readily tractable. As it is extremely difficult to empirically determine how economic agents actually make decisions, the physics-inspired theoretical analysis of toy models has a significant part to play, in particular to determine the robustness of qualitative findings to specific modeling choices. {Besides, as argued in \cite{bouchaud_crises_2013} and by some of us in \cite{garnier2022bounded}, the intrinsically individualistic nature of agent-specific moves in socioeconomic models means that the description of collective behaviors as the minimization of some global energy is often not possible. Understanding simple out-of-equilibrium dynamics as those that arise from the decision rules presented here is therefore also necessary from the methodological point of view.}

The purpose of this paper is to assess, within a general Sakoda-Schelling like occupation model, whether and how the sub-optimal concentration of agents in overly dense  regions still occurs out of equilibrium. Most importantly, we relax the assumption of taking a specific decision rule, and no longer require pre-defined block neighborhoods as in~\cite{grauwin_bertin_2009}. 
The resulting heterogeneity of interactions in our model then requires the use of out-of-equilibrium statistical mechanics techniques, the progress of which in the last decade can be credited to active matter theory. Overall, we find that the phenomenology of the model is largely unaffected by its nonequilibrium nature, suggesting that the tendency of agents to aggregate sub-optimally is robust to large classes of decision rules. This being said, our analysis highlights interesting theoretical subtleties, notably related to the non-monotonicity of the utility functions considered, that may, in turn, contribute to the understanding of other complex physical systems. 

The paper is organized as follows. In Sec.~\ref{sec:model} we introduce a  Schelling-like occupation model, in which we keep the utility function and decision rule as general as possible to allow for nonequilibrium dynamics. We the perform a numerical analysis of the model. In Sec.~\ref{sec:masterEq_LSA} we present a mean-field description of the dynamics, and determine the region in parameter space where condensation necessarily occurs. In  Sec.~\ref{sec:mappingAMB} we show how the dynamics can be mapped on the Active Model B~\cite{wittkowski2014}, which is considered to be the natural nonequilibrium extension of the Cahn-Hilliard field relaxation \cite{Cahn1958}. This mapping notably allows to compute the phase densities of the concentrated states. In Sec.~\ref{sec:extensions} we propose some relevant generalizations of the model, namely with two different populations and a housing market. Finally, in Sec.~\ref{sec:discussion} we discuss the implications of our study and conclude. 

\section{A Sakoda-Schelling occupation model}
\label{sec:model}

\subsection{Setup}
\label{subsec:model_definition}

Consider a  city structured as a two-dimensional rectangular lattice composed of $M=L_x\times L_y$ sites (or houses).  Each site can be occupied by at most one of the  $N(\leq M)$ agents living in this city. On each site of coordinate $\bm r=(i,j)$, the occupation field $n$ takes the value $n(\bm r)=1$ if the site is occupied, $n(\bm r)=0$ if it is vacant. 
It is assumed that each agent $k$ wants to maximize their own utility $u_k$, which depends on the local density of agents around them. Typically, it is natural to think that people like to gather in relatively dense areas to benefit from the city life, but not too dense as crowding might degrade the quality of life. Agents estimate the local density by averaging  the occupation field with a probability-density-function kernel $G_\sigma$, where $\sigma$ stands for the interaction range. The kernel is assumed to be isotropic and identical for all agents. The smoothed occupation field $\tilde n$ at site $\bm r$ is thus given by the discrete convolution
\begin{align}
    \tilde n(\bm r) = \sum_{\bm r'} G_\sigma(\bm r-\bm r') n(\bm r').
\end{align}

At each time step, an agent $k$ can decide to move out from their occupied site $\bm r_k$ and to settle on a new, randomly chosen, empty site $\bm r_k'$ where the utility $u[\tilde n(\bm r_k')]$ -- quantifying the agent's satisfaction -- might exceed their previous utility $u[\tilde n(\bm r_k)]$. We assume that the decision to move to the new site is a function of the utility difference $\Delta u_k\equiv u[\tilde n(\bm r_k')]-u[\tilde n(\bm r_k)]$. While the very existence of the utility function is debatable from a behavioural standpoint \cite{luce1977choice}, classical economics has traditionally taken agents to be strict utility maximizers, meaning the move will be accepted if $\Delta u_k > 0$ and rejected otherwise. In order to mitigate this assumption, a common approach is to introduce a stochastic decision rule of the form
\begin{align}
    \mathbb P(\bm r_k\to\bm r_k')=f_\rational(\Delta u_k),
\end{align}
where the function $f_\rational$ is larger than $\frac12$ whenever $\Delta u_k>0$.
Typically, $f_\rational$ is a positive and monotonic function of the utility difference, with $\lim_{x\to-\infty}f_\rational(x)=0$ and $\lim_{x\to+\infty}f_\rational(x)=1$ \cite{anderson1992discrete}.
The parameter $\rational \geq 0$, known as the \textit{intensity of choice}, or simply  the \textit{rationality}, quantifies the propensity of agents to go for strict utility maximizing. In particular, $\rational\to0$ corresponds to random decision making, while $\rational\to \infty$ means perfectly rational agents. 

In reality, the specific shape of the function $f_\rational$ is unknown. In the socio-economics literature, it is most of the time taken as the logistic function
\begin{align}
    f_\rational(x)=\frac{1}{1+e^{-\rational x}},
\end{align}
defining the so-called \emph{logit rule} \cite{luce1959individual,anderson1992discrete}. The various reasons and justifications of this decision rule are discussed and summarized in~\cite{bouchaud_crises_2013}. In a nutshell, it can be motivated axiomatically \cite{luce1959individual}, or by the fact that $f_\rational$ is a maximum entropy distribution and therefore optimizes an exploration-exploitation tradeoff when the cost associated with information scales as $1/\rational$ \cite{nadal1998formal,marsili1999multinomial}. As empirical evidence supporting this choice remains extremely scarce, its popularity is in reality largely motivated by convenience \cite{luce1977choice}. Indeed, many calculations are made possible thanks to the fact that it preserves detailed balance with respect to the Gibbs-Boltzmann measure in the particular case where agents' utility change also coincides with a global utility difference \cite{brock2001discrete}. In this context, $T \equiv 1/\rational$ can naturally be interpreted as the temperature, or ``social temperature'', of the system. In the following, the function $f_\Gamma$ will be left unspecified, unless stated otherwise. In the Monte Carlo simulations we will notably use the logit rule for simplicity.

\begin{figure}
    \centering
    \includegraphics[width=\linewidth]{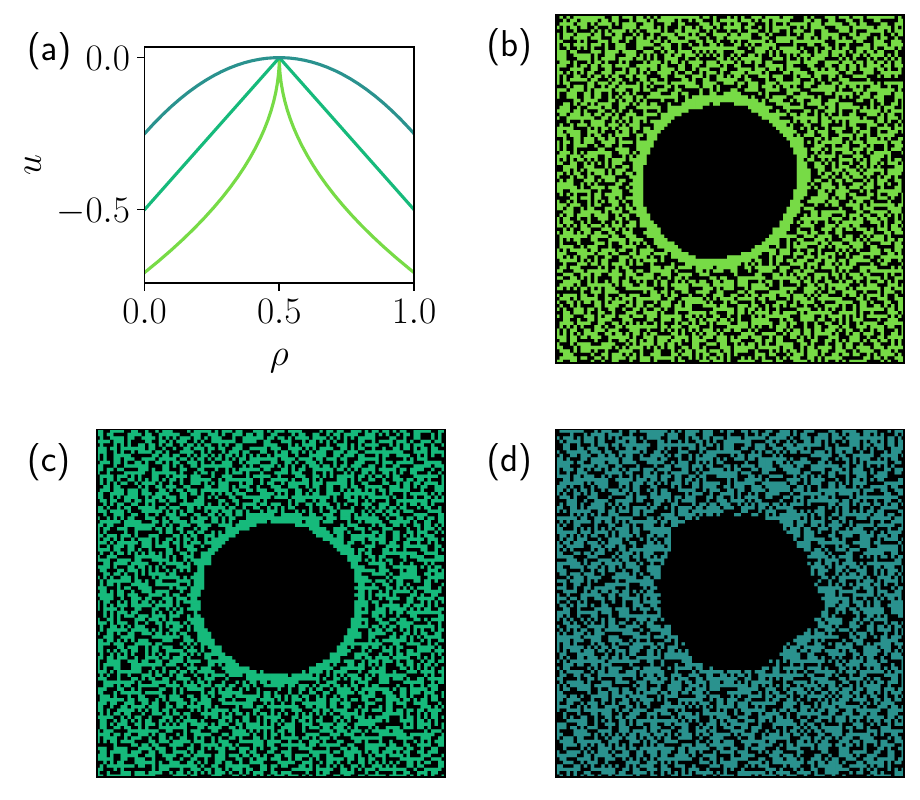}
    \caption{(a) Utility function $u(\rho)=-|\rho-\rho^\star|^\alpha$ for $\rho^\star=0.5$, $\alpha = \{0.5,1,2\}$. Panels (b), (c) and (d) show snapshots of the stationary state for these different utility functions, starting from the same homogeneous profile at $\rho_0=0.5$. Here $\rational=100$, $\sigma=3$ and $L_x=L_y=100$. The stationary density $\rho_d$ in the dense phase is $\rho_d=0.575(5)$ for $\alpha=0.5$ in (b), $\rho_d=0.575(5)$ for $\alpha=1$ in (c), and $\rho_d=0.585(5)$ for $\alpha=2$ in (d). These bulk densities are all significantly higher than the density $\rho^\star$ for which agents maximize their utility. Note the accumulation of agents at the edge of the empty domain in (b) and (c), see Sec.~\ref{sec:mappingAMB}.}
    \label{fig:utilities_snapshots}
\end{figure}

The last ingredient to specify is the utility function $u$ of the agents. As stated above, we assume that the utility depends on the locally smoothed occupation $\tilde{n}$ only, and that it is non monotonic.  As in Ref.~\cite{grauwin_bertin_2009}, we  assume that the utility is maximal for some density $\rho^\star\geq\frac12$. 
We specifically choose for the simulations
\begin{align}
    u(x)= -\left|x-\rho^\star\right|^\alpha,
\end{align}
with $\alpha>0$, see Fig.~\ref{fig:utilities_snapshots}(a), but theoretical computations below will keep $u$ unspecified.

\subsection{In or out of equilibrium?}
\label{sec:out_of_equ}

As  mentioned above, an often unspoken motivation for the use of the logit rule in the modeling of socioeconomic systems is that it may satisfy detailed balance. Indeed, as described by Grauwin \textit{et al.} \cite{grauwin_bertin_2009,grauwin2012dynamic} or in \cite{bouchaud_crises_2013} in a more general setting, if one manages to find a system-wide energy-like function $\mathcal{H}$  such that
\begin{align}
\begin{split}
    \Delta u_k &= \mathcal{H}([ \{n(\bm r)\}, n(\bm r_k')=1, n(\bm r_k)=0] )\\
    & \quad -\mathcal{H}( [\{n(\bm r)\}, n(\bm r_k')=0, n(\bm r_k)=1] ),
    \end{split}
\label{eq:linkage}
\end{align}
then the usual tools of equilibrium statistical mechanics can be used. The steady-state distribution of agents is notably identified as the minimum of the free energy, which is a Lyapunov function of the dynamics prescribed by the logit rule.

At the agent level, the existence of such a global quantity is usually the symptom of either altruistic individuals (that voluntarily maximize some collective satisfaction) or of a central planner (that constructs individual rewards towards a collective objective). Outside of these two cases, the existence of a free energy when agents are individualistic is in fact restricted to a limited number of carefully chosen models (see \cite{garnier2022bounded} for a related discussion in the context of microeconomics). In the literature of Schelling-like models, taking a city divided in neighborhoods or blocks~\cite{grauwin_bertin_2009}, where agents share the same utility, yields such a free energy description (which is importantly not a simple aggregation of individual utilities).
In our model, however, this is no longer true. 
\begin{figure}
    \centering
    \includegraphics[width=0.75\linewidth]{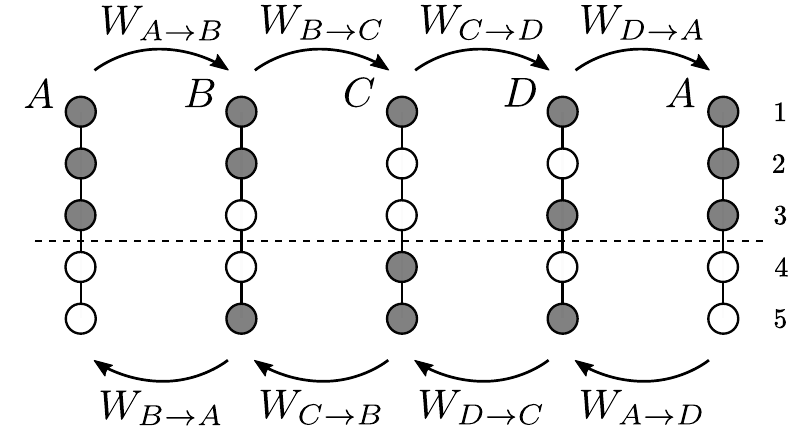}
    \caption{Loop of four configurations with $N=3$ agents on $M = 5$ sites breaking Kolmogorov's criterion when the utility is non-linear and is a function of an individual perceived density. Shaded and unshaded nodes correspond to occupied and empty sites respectively. The dashed line indicates a possible segmentation of the system into two distinct neighborhoods.}
    \label{fig:Kolmogorov}
\end{figure}

To explicitly show that the dynamics breaks detailed balance despite the logit rule, one may consider a small system and find a specific cycle breaking Kolmogorov's criterion \cite{kolmogoroff1936theorie}. Such a cycle between four consecutive states with $N = 3$ agents placed on a one-dimensional ``city'' with $M = 5$ sites is illustrated in Fig.~\ref{fig:Kolmogorov}. The ratio of transition rates between states $X$ and $Y$, that differ by an agent located on sites $i$ in $X$, versus $j$ in state $Y$, is given by
\begin{equation}
\frac{W_{X\to Y}}{W_{Y\to X}} = \frac{1+\ee^{-\rational [u(\tilde{n}_i^X) - u(\tilde{n}_j^Y)]}}{1+\ee^{-\rational [ u(\tilde{n}_j^Y) - u(\tilde{n}_i^X)]}} = \ee^{\rational [ u(\tilde{n}_j^Y) - u(\tilde{n}_i^X)]}.
\end{equation}
As a result, the ratio between the product of forward rates, $W_+$, and the product of backwards rates, $W_-$, in the cycle shown in Fig.~\ref{fig:Kolmogorov}, is given by
\begin{equation}
\begin{aligned}
    \frac{W_+}{W_-} = \ee^{\rational [u(\tilde{n}_5^B) - u(\tilde{n}_3^A) - u(\tilde{n}_2^B) + u(\tilde{n}_3^D) + u(\tilde{n}_2^A) - u(\tilde{n}_5^D)]}.
    \label{eq:rates_ratio}
\end{aligned}
\end{equation}
For a generic non-linear utility function, $W_+ \neq W_-$, which is a signature of  nonequilibrium dynamics. For a linear utility function on the other hand, considering that the convolution kernel $G_\sigma$ is isotropic, all terms in the exponential cancel out, leading to $W_+=W_-$ (which would be also satisfied for any other cycle). In this situation, the utility difference can simply be interpreted as an energy difference, where the kernel $G_\sigma$ plays the role of a pairwise interaction potential between the agents. Interestingly, this small cycle also illustrates how the introduction of neighborhoods can salvage the equilibrium description for a generic utility. Splitting the lattice in two neighborhoods along the dashed line shown in Fig.~\ref{fig:Kolmogorov} and taking an identical value of $\tilde{n}$ for all agents on each neighborhood, the terms in the exponential in Eq.~\eqref{eq:rates_ratio} indeed cancel out for any utility function since $\tilde{n}_5^B = \tilde{n}_5^D$, $\tilde{n}_3^A = \tilde{n}_2^A$ and $\tilde{n}_2^B = \tilde{n}_3^D$.

\begin{figure*}[t]
    \includegraphics[width=\linewidth]{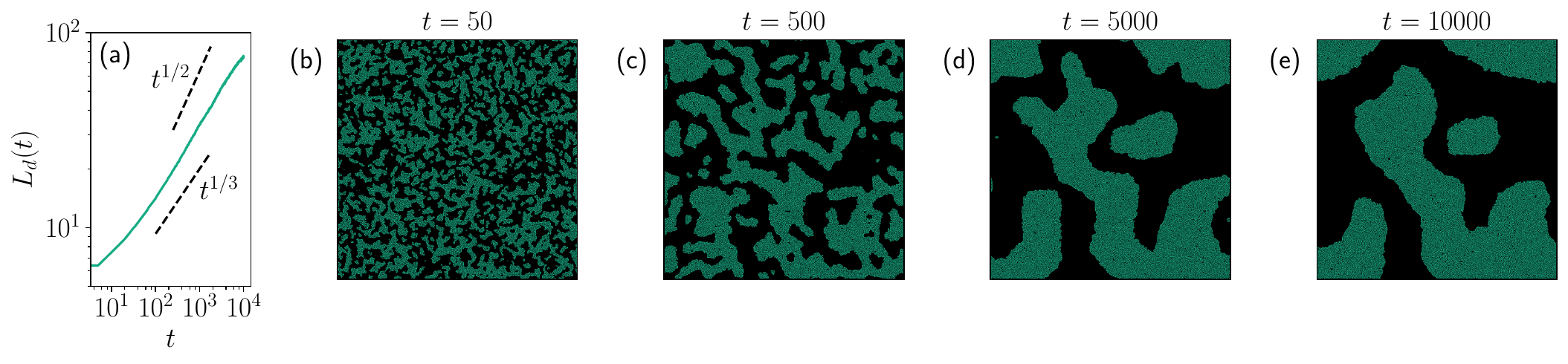}
    \caption{(a) Typical dense domain size $L_d(t)$ during coarsening as a function of time $t$. A unit of time is defined as $N$ Monte Carlo steps, where $N$ is the number of agents. $L_d(t)$ is averaged over 5 independent simulations. (b), (c), (d) and (e) show snapshots at different times. Starting from a disordered configuration, we quench the system at low temperature, or high rationality $\rational$, corresponding to $T\simeq T_c/6$. Parameters: $L_x=L_y=600$, $\rho_0=0.3$, $\sigma=1$, $\alpha=3/2$, $T = 0.01$.}
\label{fig:coarsening_exponent_snapshots}
\end{figure*}

\subsection{Microscopic simulations}
\label{subsec:micro_simulations}

Having established the out-of-equilibrium nature of our model, we start by performing numerical simulations to assess whether the concentration of agents in overly dense regions is generic and  robust to different shapes of the utility function. Here, all numerical simulations are performed on a two-dimensional grid with periodic boundary conditions. The utility is maximal for $\rho^\star=1/2$. For the sake a simplicity, here we use the logit decision rule and a truncated Gaussian kernel
\begin{equation}
    G_\sigma(\bm r) = \begin{cases} 
    \frac{1}{N_\sigma} \ee^{-\frac{1}{2\sigma^2} \| \bm r\|^2},\quad \text{ if $\| \bm r\|$ }\leq 4\sigma,\\
    0, \quad \text{otherwise},
    \end{cases}
\end{equation}
 where $N_\sigma$ enforces the normalization of the kernel.

\subsubsection{Phase separation}
For large system size $L_x,L_y\gg \sigma$, we explore the behavior for different  global densities $\rho_0=N/(L_xL_y)$ and for various rationality parameters $\rational$. Numerical results are qualitatively similar for all the values of $\alpha$ we tested, ranging from $\alpha=0.5$ to $\alpha=2$, see Fig.~\ref{fig:utilities_snapshots}. The phenomenology can be summarized as follows. When rationality is low ($\rational\to 0$, $T\to\infty$), the stationary state remains homogeneous because agents settle at random. When rationality is high, agents may aggregate in dense clusters, which can surprisingly be more crowded than what agents' utilities prescribe. This  was already discussed in \cite{grauwin_bertin_2009} where the authors point out that the homogeneous state is actually an unstable Nash equilibrium, even though all agents maximize their utility. The destabilization occurs as one agent randomly moves to another region (with no regard to the effect it may have on the other agents utilities), which decreases the average density at their original site and increases the average density where they settle. Agents in the lower-density region will eventually move to gain the utility they lost when their neighbors moved out. This dynamics will eventually empty some regions, in which agent's return becomes statistically less and less probable. The final state, where a dense phase and an empty phase coexist, is a stable Nash equilibrium. 

One can quantify the condensation dynamics when starting from the homogeneous state and taking high rationality. The system undergoes a spinodal decomposition where dense clusters grow and merge until there is one large dense cluster only, as shown in Fig.~\ref{fig:coarsening_exponent_snapshots}. The final cluster topology ultimately depends on noise realization and on the box dimensions. We measure the cluster size $L_d(t)$ as a function of time $t$ using the radial structure factor (see App.~\ref{app:coarsening}). We find $L_d(t)\sim t^{1/z}$, with the dynamical exponent $z\in[2,3]$, reminiscent of the coarsening exponent observed in a 2D Ising system with long-range Kawasaki dynamics~\cite{tamayo_critical_1989, bray_comment_1991, rutenberg_nonequilibrium_1996, tartaglia_coarsening_2018}. Interestingly, and consistent with the findings of \cite{tartaglia_coarsening_2018} in the low temperature region, our results suggest an exponent closer to the local Kawasaki dynamics result $z = 3$ (see Fig.~\ref{fig:coarsening_exponent_snapshots}(a)), despite long-range particle displacements.

\subsubsection{Critical point and critical exponents}

The complete phase separation that occurs when  rationality is high indicates the use of the order parameter $m\equiv \rho_d -\rho_g$, where $\rho_d,\rho_g$ are the average densities of the dense and ``gas'' (dilute) phases, respectively. At the critical point $(\rho_c,T_c)$, we expect a second-order phase transition where $m$ goes to $0$ with power-law scaling
\begin{align}
     m  \underset{\tau\to 0^+}{\sim} \tau^{\beta},
\end{align}
where $\tau=(T_c-T)/T_c>0$ defines the rescaled temperature difference, and $\beta$ is the order-parameter critical exponent. Measuring the critical exponents allows one to determine to which universality class the system belongs to, providing precious information on the system behavior at large scales. 
Since simulations are carried out in finite systems, measuring the critical point with precision requires numerical ruse. We follow the approach that has been extensively used to measure critical exponents in systems undergoing a Motility-Induced Phase Separation (referred to as MIPS) \cite{siebert_critical_2018, maggi_universality_2021, dittrich_critical_2021}, see App.~\ref{app:critical_exponents_add}.

\begin{figure}
\includegraphics[width=\linewidth]{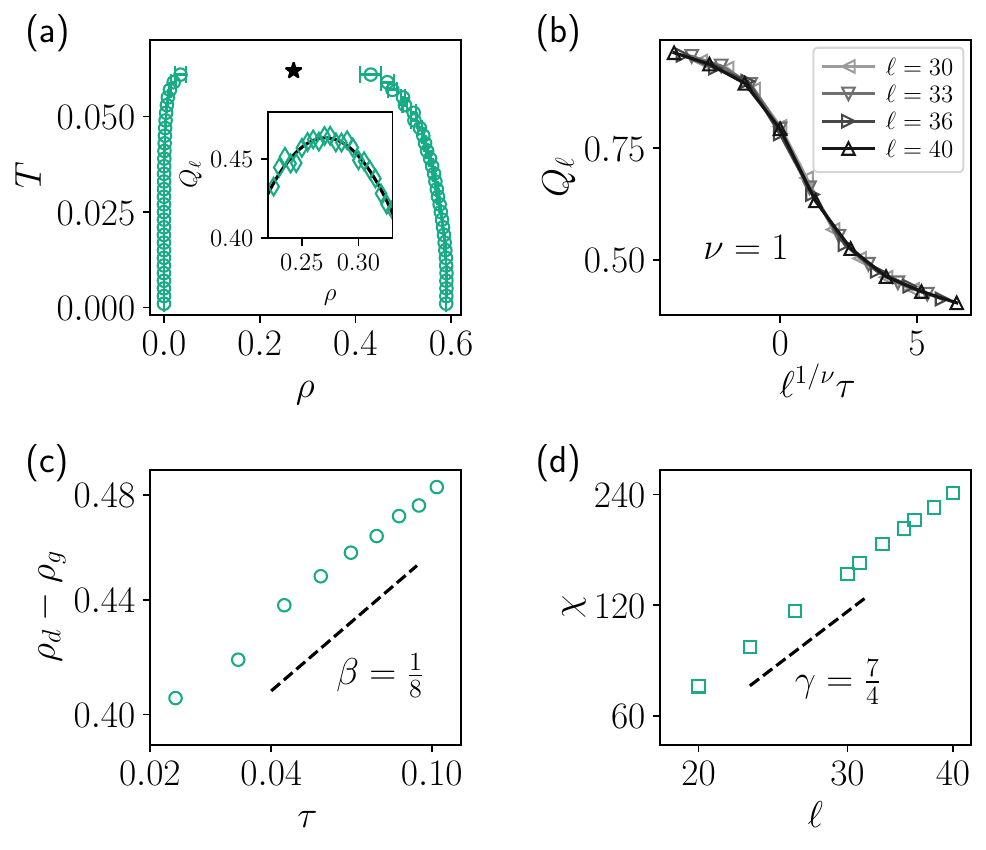}
    \caption{Numerical experiments for $\sigma=1$, $\alpha=3/2$. (a) Binodal densities measured for $L_x = 200$ and $L_y = 66$ ($\ell = 33$), inset showing the Binder cumulant as a function of the density and fitted (continuous line) to determine the critical density. (b), (c) and (d) show the numerical measurements of the critical exponents close to the critical point $(\rho_c,T_c)=(0.271,0.0620)$ determined using various system sizes ranging from $\ell = 20$ to $\ell = 40$.}
    \label{fig:phaseDiag_criticalExp}
\end{figure}

Simulations are performed in a rectangular domain of size $L_x\times L_y$, with $L_x=3L_y$, with periodic boundary conditions to keep flat interfaces between a stripe of liquid (dense phase) and a stripe of gas (dilute phase).  Starting with the dense phase in the center of the system, we track the center of mass such that we always compute the densities in the bulk of each phase. 
To compute the local density inside the bulk of each phase, we consider square boxes of size $\ell=L_y/2$, centered either in $0$ in the gas bulk or centered in $L_x/2$ in the dense bulk (Fig.~\ref{fig:boxes_centered}). The local density in each box fluctuates and it is given by
$\rho_b = N_b/\ell^2$ with $N_b$ the number of agents in the box $b$ in a given realization of the system. The distribution of the density in the system is thus bimodal for $T<T_c$ and unimodal when the system is homogeneous. 
Defining 
\begin{align}
    \Delta \rho = \frac{N_b-\langle N_b \rangle}{\ell^2},
\end{align}
where the $\langle \cdot \rangle$ stands for averaging on the four boxes and on independent realizations of the simulation,
we compute the celebrated Binder cumulant~\cite{rovere1990, *rovere_simulation_1993,binder1997} 
\begin{align}
    Q_\ell(\Delta \rho, T) = \frac{\langle (\Delta \rho)^2\rangle}{\langle (\Delta \rho)^4\rangle},
\end{align}
for a given box size $\ell$ and a given temperature $T$.
For $\ell$ large enough, the curves $Q_\ell(T)$ all intersect in $T=T_c$ where the behavior of the system is universal. It is important to mention that the critical density is not known \emph{a priori}. It has to be assessed beforehand to ensure that the system, as $T$ changes, goes through the critical point, where the phase transition is of second-order type. To locate $\rho_c$, we compute the Binder cumulant at fixed temperature, close to the estimated critical point, for various densities $\rho_0$. The critical density then corresponds to the maximal fluctuations of $\Delta \rho$, translated in a peak of the Binder cumulant, see Fig.~\ref{fig:phaseDiag_criticalExp}(a).
Once the critical point is precisely located, additional critical exponents can be measured. Notably, defining the susceptibility $\chi$ as
\begin{align}
    \chi\equiv \frac{\langle(N_b-\langle N_b\rangle)^2\rangle}{\langle N_b \rangle}= \frac{\langle (\Delta \rho)^2\rangle}{\langle N_b\rangle} \ell^4,
\end{align}
one obtains
\begin{align}
    \chi \sim \ell^{\gamma/\nu},\quad
    \frac{d Q_\ell }{d\tau}\Big|_{\tau=0}\sim \ell^{1/\nu},
\end{align}
at the critical point. 

We report in Fig.~\ref{fig:phaseDiag_criticalExp} the various results on the critical point and on the critical exponent for $\sigma=1$ and $\alpha=3/2$. Using the Binder cumulant, one identifies the critical point at $\rho_c=0.271(5)$ and $T_c=0.0620(2)$, where the uncertainty on the last digit appears in the parentheses. The phase diagram in space $(\rho,T)$ is shown in Fig.~\ref{fig:phaseDiag_criticalExp}(a), the black star indicates the critical point and the circular markers show the densities of the coexisting phases: they  define the \emph{binodal} frontier. The exponent $\beta$ is directly measured from the order parameter $m$ as function of reduced temperature $\tau$, at a fixed system size $L_x=220$ [see Fig.\ref{fig:Binder_collapse}(c)]. From the Ising-2D ansatz, we check that $\nu=1$ yields a neat collapse of the Binder cumulant, see Fig.~\ref{fig:Binder_collapse}(b). The exponent $\gamma$ is obtained by varying the system size at the critical temperature $T_c$ and assuming $\nu=1$ [see Fig.\ref{fig:Binder_collapse}(d)]. We report in Table~\ref{tab:report_exponent} the values found for the critical exponents in the cases $\alpha = 3/2$ (Fig.~\ref{fig:phaseDiag_criticalExp}) and $\alpha = 1/2$ (not shown here). 
\begin{table}[]
    \centering
    \begin{tabular}{c|c|c|c|c}
    Model & $\rho_c$ & $T_c$ & $\beta$ & $\gamma$   \\
   \hline
   \hline
    Ising 2D (exact)  & 0.5 & &  0.125  & 1.75\\
    $\alpha=1/2$, $\sigma=1$ & 0.309(5) & 0.0983(5) & 0.120(8) & 1.71(5)\\
    $\alpha=3/2$, $\sigma=1$ & 0.271(5) & 0.0620(2) & 0.119(5) & 1.74(5)
    \end{tabular}
    \caption{Critical density and exponents for nonequilibrium Sakoda-Schelling model for $\alpha=1/2$ and $\alpha=3/2$.}
    \label{tab:report_exponent}
\end{table}
They differ by less than 5\% from the 2D Ising static exponents. These results enjoin us to assert with a high degree of confidence that the model considered here belongs to the 2D-Ising universality class. Since the system is out of equilibrium and particle displacements can be of infinite range, recovering the Ising universality class is \textit{a priori} nontrivial. However, finding other critical exponents would have been surprising since the ingredients at play are the ones of the Ising model, namely, short-range and isotropic interactions, a homogeneous medium and a two state degree of freedom (sites are empty or occupied).
This result must also be put into perspective with the recent debate on the universality class(es) of systems undergoing MIPS~\cite{siebert_critical_2018,dittrich_critical_2021,maggi_universality_2021, gnan_critical_2022,nakano2023universal}, and their associated active field theories~\cite{caballero_strong_2018,speck_critical_2022}. Notably here, our interaction kernel $G_\sigma$ provides a so-called \emph{quorum-sensing} interaction, like that found in assemblies of bacteria~\cite{curatolo_cooperative_2020}. The particle dynamics is however quite different for bacteria and for our agents. The remaining of the paper shall be devoted to establishing a quantitative relation between our Sakoda-Schelling occupation  model and the field-theory descriptions of MIPS. The first step along this path is to formulate a mean-field approximation of our model.

\section{Field theory and the local-move approximation}
\label{sec:masterEq_LSA}

\subsection{General description}
The computation starts by writing the expectation of the occupation number $n_{\bm r,s+1}\equiv n(\bm r,s+1)$ of site $\bm r$ at time $s+1$, conditioned on the 
previous configuration $\{n_{\bm r,s}\}$.
Averaging over multiple realizations of noise and using a mean-field approximation in which all correlation functions factorize, one obtains
\begin{align}
\begin{split}
    \langle n_{\bmr,s+1}\rangle  - \langle n_{\bmr,s}\rangle  &= (1-\langle n_{\bmr,s}\rangle)\sum_{\bmr'\neq \bm r}\langle n_{\bmr',s}\rangle f_\rational(\Delta u^s_{\bmr'\to \bmr})\\
    &- \langle n_{\bmr,s}\rangle\sum_{\bm r'\neq \bm r} (1-\langle n_{\bmr',s}\rangle) f_\rational(\Delta u^s_{\bmr\to \bmr'}),
    \label{eq:MF_occupation_number}
\end{split}
\end{align}
where $\Delta u^s_{\bmr \to \bmr'}\equiv u(\langle \tilde n_{\bm r',s}\rangle)-u(\langle\tilde n_{\bm r,s}\rangle)$. 
For convenience, we take the continuous time and continuous space limit, following the common procedure to obtain a mean-field description of  exclusion processes on lattices (see e.g. \cite{curatolo_multilane_2016}). 
The average occupation number $\langle n\rangle$ is now described by the density $\rho$, while the spatially smoothed average occupation number $\langle \tilde n\rangle$ is described by the field $\phi\equiv G_\sigma*\rho$.
The master equation for the occupation probability then takes the form of a noiseless hydrodynamic equation, in our case:
\begin{equation}
\begin{aligned}
    \partial_t\rho(x,t) =& [1-\rho(x,t)] \int \dd y \, \rho(y,t)w_\rational([\phi],y,x,t)  \\
    &- \rho(x,t) \int \dd y \, [1-\rho(y,t)] w_\rational([\phi],x,y,t),
\end{aligned}
\label{eq:MF_hydro}
\end{equation}
with the transition rate from $y$ to $x$ explicitly given by
\begin{align}
    w_\rational([\phi],y,x,t)= \omega f_\rational\left(u(\phi(x,t))-u(\phi(y,t))\right), 
    \label{eq:logit_rate}
\end{align}
where $\omega$ is homogeneous to an inverse time scale and $f_\rational$ is left unspecified. Equation~\eqref{eq:MF_hydro} is valid in any dimension, but, for simplicity, we will work out the mean-field computations in dimension 1 in space. This can be justified \emph{a posteriori} when we compare the mean-field (MF) to the Monte Carlo (MC) simulations. Let us also mention that the dimension does not play a role in determining the phase densities in the steady state of coarse-grained field theories (Allen-Cahn~\cite{allen1979microscopic}, Cahn-Hilliard~\cite{cahn_hilliard}, etc.).

Integrating Eq.~\eqref{eq:MF_hydro} over space, one immediately sees that the total density $\int \rho$ is conserved. One can also check that in the very specific case where $u(\phi)$ is linear in $\phi$, 
one can build a free-energy functional that is a Lyapunov function of the non-local MF dynamics, ensuring a convergence towards local minima and preventing limit cycles and oscillatory dynamics.
This is a natural consequence of the fact that detailed balance is satisfied at microscopic level.
In App.~\ref{app:lyapunov_nonlocal}, we construct this free energy and  show that the dynamics is relaxational. 

\subsection{Linear stability analysis}

In the general case, we would like to understand how the homogeneous state becomes unstable. To do so, we consider a small perturbation around the homogeneous state: 
$\rho(x,t)=\rho_0 + \rho_1(x,t)$, with $\rho_1$ the perturbation. By linearity of the convolution, one has  $\phi(x,t) = \rho_0+\phi_1(x,t)$, with $\phi_1\equiv G_\sigma*\rho_1$. A Taylor expansion of Eq.~\eqref{eq:MF_hydro} combined with mass conservation (i.e $\int_D \rho_1=\int_D\phi_1=0$, where $D$ is the full domain),  finally yields:
\begin{align}
\begin{split}   
\partial_t\rho_1(x,t)=& \ 2\Omega\rho_0(1-\rho_0)f'_\rational(0)u'(\rho_0)\phi_1(x,t)\\ &-\Omega f_\rational(0)\rho_1(x,t),
\end{split}
\end{align}
with $\Omega$ the full domain size.
Defining the Fourier transform for any field $h$ as $\hat h(k) = \int \dd x \, e^{-i k x} h(x)$, one obtains 
\begin{align}
&\partial_t \hat \rho_1(k,t) =\Lambda(k)\hat\rho_1(k,t),\\
    &\Lambda(k)= \Omega f_\rational(0)\left(2\rho_0(1-\rho_0)\frac{f'_\rational(0)}{f_\rational(0)}u'(\rho_0)\hat G_\sigma(k)- 1\right).
\end{align}
This last equation shows that the homogeneous state is unstable if there exists a mode $k^\star$ such that 
\begin{align}
    2\rho_0(1-\rho_0)\frac{f'_\rational(0)}{f_\rational(0)}u'(\rho_0)\hat G_\sigma(k^\star)>1.
    \label{eq:LSA_threshold_full}
\end{align}
The manifold for which the inequality becomes an equality defines the spinodal in the phase diagram $(\rho_0,\rational)$.
In particular, for any monotonically decreasing kernel $G_\sigma(|x|)\in L_2(\mathbb{R})$, one has $\hat G_\sigma(0)>|\hat G_\sigma(k)|$, such that for large system size, the stability of the homogeneous state is given by the stability of modes $k\to0$, and the spinodal is thus defined by the equation
\begin{align}
   2 \rho_0(1-\rho_0)\frac{f'_\rational(0)}{f_\rational(0)}u'(\rho_0)=1.
    \label{eq:LSA_general}
\end{align} 
Note that this criterion is generic as it only depends on the
 decision rule through $f_\Gamma(0)$ and $f'_\Gamma(0)$.
The simulations also reveal the existence of a bistable region in the vicinity of this spinodal. This is the binodal region, where hysteresis and bistability can notably be observed, and which can be fully characterized in the case of an equilibrium system~\cite{gauvin2009_phaseDiag}. Here however, there is a priori no free energy one can rely on to describe the nucleation scenario and to obtain the densities of the phase-separated state. 

\begin{figure}
 \centering
 \includegraphics[width=\linewidth]{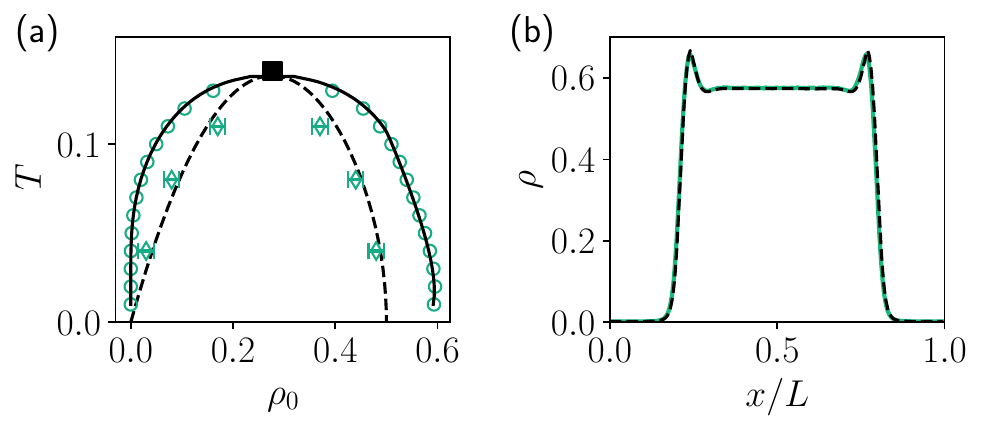}
  \caption{Comparison between Monte Carlo simulations and mean-field results for $\alpha = 3/2$, $\sigma=7$ and $L_x=200$, $L_y = 66$ ($\ell = 33$).  
  (a) Phase diagram in the $(\rho_0,T)$ plane. The mean-field binodal (continuous black line) is given by measuring the densities of the bulk of each plateau in a phase separated state. The circles are the bulk averaged densities in Monte Carlo (MC) simulations. The dashed black line represent the mean-field spinodal, which is obtained analytically from the linear stability analysis (see Eq.~\eqref{eq:LSA_threshold_full}), with $\hat G_\sigma(k_1)=e^{-\sigma^2 k_1^2/2}$ and $k_1=2\pi/L_x$. The diamonds indicate the loss of stability of the homogeneous state in the MC simulations. The black square is the critical point for $\sigma/L\to 0$. 
  (b) Averaged density profile $\rho(x)$ from MC simulations (continuous green line) for $\rho_0=0.35$, $T=0.05$. The dashed black line is the stationary solution of the mean-field equation Eq.~\eqref{eq:MF_hydro} for the same parameters, solved on a grid of step size 1 with a Euler explicit scheme.
  }
  \label{fig:compare_profile_phaseDiag}
\end{figure}

\subsection{Comparison to microscopic simulations}

The MF prediction is expected to be accurate for systems with high connectivity, which here corresponds to large $\sigma$. In the following, we shall take the limit $L\to \infty$, $\sigma \to \infty$ with $\sigma/L \to 0$ to obtain mean-field predictions that are independent of both $\sigma$ and $L$, and perform numerical simulations as close as possible to this scaling regime.

The first analytical prediction of the MF description is the spinodal, that determines the onset of instability of the homogeneous state, see Eq.~\eqref{eq:LSA_general}. The spinodal is the dashed line in the $(\rho,T)$ phase diagram in Fig.~\ref{fig:compare_profile_phaseDiag}(a). To check the prediction, we start in the MC simulations from a uniformly distributed configuration of agents for three different values of temperature, $T=0.04$, $0.08$, $0.11$, and we detect the frontier across which the homogeneous profile either coarsens, or needs a nucleation event to converge to the separated state. This frontier is marked with the diamonds, which agrees with the MF prediction.

Second, the MF dynamical Eq.~\eqref{eq:MF_hydro} can be solved numerically with an Euler explicit scheme. From the numerical solution, one obtains the densities of the bulk of each phase when a phase separation occurs: these densities define the binodal, the continuous line in Fig.~\ref{fig:compare_profile_phaseDiag}(a). These MF phase densities are perfectly recovered by the MC simulations (circles). 
In addition, one can compare the steady-state average density profile from MC simulations to the mean-field stationary density, which superimpose almost exactly, see Fig.~\ref{fig:compare_profile_phaseDiag}(b).

As previously stated, the MF predictions fail for small values of $\sigma$. The phase diagram in Fig.~\ref{fig:phaseDiag_criticalExp}(a) is for instance obtained for $\sigma=1$, and indeed strongly differs from the MF solution. For $\sigma=1$, we notably identify the critical point at $(\rho_c,T_c)=(0.271, 0.0620)$, whereas the MF predicts $(\rho_c,T_c)_\mathrm{MF}=(0.2763, 0.1418)$, where, as expected, $T_c^{\sigma=1} < T_c^{\mathrm{MF}}$.

\subsection{Local-move approximation}

To make progress into the identification of a possible effective free energy functional, it may be convenient to consider slightly modified dynamics where jumps are now only authorized in the direct neighborhood of the agents. Indeed, considering an evolution enforcing a \textit{local} mass conservation will allow for more familiar partial differential equations (PDEs) and field theoretic approaches on conserved scalar fields. Here, the absence of macroscopic density currents in the steady state, both in MC simulations and in the MF solution suggests that the system generically converges to a stationary stable fixed point, where the details of the dynamics become inconsequential.  In addition, when the majority of agents have aggregated in a single dense cluster in the steady state,  it is unlikely that they would perform moves outside of the bulk, in low-density regions, since the utility there is minimal. The local-move approximation, as it strongly simplifies the description, thus appears natural.\footnote{Dynamically, the coarsening exponent $z\simeq 3$ displayed in Fig.~\ref{fig:coarsening_exponent_snapshots}(a), and which is also observed in a Cahn-Hilliard relaxation dynamics can also be invoked to support the idea of local moves.}

Following the Taylor expansion outlined in App.~\ref{app:LSA_localMoves}, the local mean-field dynamics is given by
\begin{align}
    \partial_t \rho &= f_\rational(0) \partial_x^2 \rho -2f_\rational'(0) \partial_x[\rho(1-\rho)\partial_x u],
\end{align}
which can be rewritten as the canonical equation for the mass-conserving dynamics
\begin{equation}
    \partial_t \rho =\partial_x[M[\rho] \partial_x \mu([\rho],x)],
    \label{eq:fermion_meanfield_conserved_1d}
\end{equation}
with the mobility operator $M[\rho] = \rho(1-\rho)$, stemming from the non-overlaping nature of the agents, and with the chemical potential $\mu=\mu_\mathrm{ent.}+\mu_\mathrm{util.}$ where
\begin{align}
    \mu_\mathrm{ent.} &=f_\rational(0)\log\left(\frac{\rho}{1-\rho} \right)\label{eq:mu_entropy}\\
    \mu_\mathrm{util.}&=-2f_\rational'(0)u[\phi(x)].
    \label{eq:mu_utility}
\end{align}
The first contribution to the chemical potential $\mu_\mathrm{ent.}$ is  purely local and accounts for entropy in the system where agents cannot overlap. The second contribution $\mu_\mathrm{util.}$ encodes the drive from agents' utility. This term exhibits non-locality with respect to the field $\rho$, and as a consequence, cannot be expressed as a functional derivative of any free energy, in general \cite{grafke2017,obyrne2020, obyrne_nonequilibrium_2023}. However, in the particular case of a linear utility in $\phi$, one again recovers that $\mu_\mathrm{util.}+\mu_\mathrm{ent.}$ can be written as the functional derivative of the free energy $\mathcal F$ given in App.~\ref{app:lyapunov_nonlocal} and, as a consequence, the dynamics \eqref{eq:fermion_meanfield_conserved_1d} becomes a gradient descent~\cite{otto_geometry_2001}. Let us emphasize that, here again, the decision rule is kept general, and that the entire local dynamics only depend on it through $f_\Gamma(0)$ and $f'_\Gamma(0)$.

Performing the linear stability analysis on the dynamics with local moves (see App.~\ref{app:LSA_localMoves}), we find that the criterion for the homogeneous solution to be unstable is identical to that given in Eq.~\eqref{eq:LSA_general}, when moves are global. Also, the stationary density profiles computed either with the local, or with the non-local MF PDEs for the same parameters are identical, as shown in App.~\ref{app:local_v_nonlocal}. Both these observations therefore allow us to confirm the relevance of the local-move approximation to characterize the system in the long-time limit.

Finally, note that the local hydrodynamic equations can also be obtained using the path integral approach on a lattice~\cite{lefevre2007dynamics}, which, in passing, provides the fluctuating hydrodynamics:
\begin{align}
    \partial_t \rho =\partial_x \left[\rho(1-\rho)\partial_x \mu([\rho],x) + \sqrt{\rho(1-\rho)} \xi\right],
\label{eq:fermion_fluctuating_conserved_1d}
\end{align}
where $\xi(x,t)$ is a Gaussian white noise with zero mean and with $\langle \xi(x,t) \xi(x',t')\rangle = 2 f_\rational(0) \delta(t-t') \delta
(x-x')$. We then remark that when the utility is linear, the stochastic field evolution describes a complete equilibrium dynamics, irrespective of the choice of the decision rule: A rule that breaks detailed balance at the microscopic level can still lead to an equilibrium field theory after coarse graining. Similar findings had been pinpointed in active matter models~\cite{tailleur_cates2008, obyrne2020}.
While not central to the present work, the fluctuations can be studied in more detail, providing information on the nucleation scenarii and on transition paths between macroscopic states for instance~\cite{tailleur_mapping_2008, bertini_macroscopic_2015, baek_dynamical_2018, zakine2022minimum}. The study of the associated functional Fokker-Planck equation using the tools described in \cite{wu2013potential,wu2014potential} may also be an interesting perspective for future works. In the case of non-local moves, the formalism from~\cite{lefevre2007dynamics} cannot be straightforwardly adapted, since the local gradient expansion of the jump rates in the action breaks down. Establishing an appropriate fluctuating hydrodynamic description in the case of non-local dynamics is therefore an open problem.

\section{Generalized-thermodynamic construction}
\label{sec:mappingAMB}

Since the previous section has shown that the phase separation is well described by the local-move approximation, we can now use the machinery of field theory for scalar active matter (e.g. Active Model B), as developed in~\cite{wittkowski2014,solonPRE2018,cates_active_2019}. This mapping will notably allow us to obtain the binodal densities for some class of utility functions detailed below.

\subsection{The generalized-thermodynamic expansion}

Even though $\mu$ in Eq.~\eqref{eq:fermion_meanfield_conserved_1d} cannot be written as the functional derivative~\cite{obyrne2020, grafke2017, obyrne_nonequilibrium_2023}, the dynamics can  be analyzed by resorting to a gradient expansion. Indeed, expanding the chemical potential up to $O(\nabla^4, \rho^2)$ terms yields
\begin{align}
    \mu[\rho] = g_0(\rho) + \lambda(\rho) (\nabla \rho)^2 -\kappa(\rho)\nabla^2 \rho +O(\nabla^4, \rho^2),
    \label{eq:expansion_mu_AMB}
\end{align}
with $g_0$, $\lambda$, $\kappa$ local function of the field $\rho$,
and a generalized thermodynamic mapping~\cite{solonPRE2018,solon_generalized_2018} can yield the prediction of the binodal densities.

For simplicity, we will now assume that the smoothing kernel is a Gaussian distribution of zero mean and variance~$\sigma^2$. In Fourier space, the smoothed field is given by $\hat\phi(k)=\hat \rho_k \exp({-{\sigma^2 k^2}/{2}})$, which can be truncated to leading order:
\begin{align}
   \hat \phi_k \simeq \hat \rho_k \left(1-\frac{k^2\sigma^2}{2}+O(\sigma^4 |k|^4 )\right).
\end{align}
In real space, this translates into $\phi=\rho+\frac{\sigma^2}{2}\nabla^2\rho + O(\nabla^4,\rho)$. This allows us to further expand the $\mu_\mathrm{util.}$ given in Eq.~\eqref{eq:mu_utility}. To leading order in the $O(\nabla,\rho)$ expansion, one has
\begin{align}
    \mu_\mathrm{util.}=-2f'_\rational(0)\left[u(\rho)+\frac{\sigma^2}{2}u'(\rho)\partial_x^2\rho + O(\partial_x^4,\rho)\right].
\end{align}
Combining this expansion of $\mu_\mathrm{util.}$ with the entropic contribution $\mu_\mathrm{ent.}$, it is now possible to identify the different terms in Eq.~\eqref{eq:expansion_mu_AMB}, namely:
\begin{align}
    &g_0(\rho)= -2f'_\rational(0)u(\rho)+ f_\rational(0)\log\left( \frac{\rho}{1-\rho}\right);\label{eq:g0}\\
    &\lambda(\rho)=0; \quad \kappa(\rho)=f'_\rational(0)\sigma^2u'(\rho).
\end{align}
This identification enables us to follow up to the next step, which is finding the proper function $R(\rho)$ and the generalized functional $\mathcal G[R]$ by means of which the dynamics will be given by
\begin{align}
    \partial_t\rho(x,t) = \partial_x \cdot \left[M[\rho]\partial_x \frac{\delta \mathcal G}{\delta R(x,t)}\Big|_{R(\rho)}\right].
\end{align}
A double-tangent construction on $\mathcal G[R]$ then provides the binodal densities~\cite{solonPRE2018}. Since $\lambda(\rho)=0$, the differential equation that the function $R$ must satisfy (see~\cite{solonPRE2018,solon_generalized_2018}) is 
\begin{align}
    \kappa(\rho) R''(\rho)=-\kappa'(\rho)R'(\rho),
\end{align}
which simplifies into $(\kappa R')'=0$,  where the $'$ denotes the derivative with respect to $\rho$. 

\subsection{Implications of non-monotonous utilities}

The previous equation suggests $R'(\rho)=C/\kappa(\rho)$, with $C$ some constant. However, one has to be careful at this stage. In the case considered here, where the utility of agents reaches its maximum for some density $\rho^\star$, it is clear that $\kappa(\rho)$ undergoes a sign change at $\rho^\star$, and more precisely since $f'_\rational(0) >0$, we have $\mathrm{sign}[\kappa(\rho)]=\mathrm{sign}[u'(\rho)]$. To our knowledge, the fact that $\kappa(\rho)$ may not remain strictly positive has never been considered in the active matter literature, even though it bears important physical meaning. Consider a system of quorum-sensing moving bacteria whose microscopic velocity $v(\rho)$ is density dependent~\cite{curatolo_cooperative_2020}. Coarse-graining typically yields $\kappa(\rho)\simeq -v'(\rho)/v(\rho)$~\cite{cates_arrested_2010,solon_generalized_2018}, and one obtains $\kappa(\rho)>0$ when the velocity of the particles is a decreasing function of local density, eventually leading to bacteria condensation, i.e. MIPS ~\cite{tailleur_cates2008, cates_tailleur_MIPS2015}. A positive $\kappa>0$ is thus naturally interpreted as an ``effective surface tension'' in this framework. On the other hand, a negative $\kappa(\rho)$ would be the reflect of an increasing motility with increasing bacterial density, which is also biologically relevant if one considers that bacteria are likely to avoid competition for resources in crowded areas. 
Yet, and this is a key remark here, it does not necessarily mean that the phase separation is arrested or that the system undergoes a microphase separation when $\kappa<0$, notably because higher order gradient terms that were discarded in the field expansion then become relevant and may stabilize the interfaces. More specifically here, for $\rho>\rho^\star$, $u'(\rho)<0$ such that the term $O(\partial_x^2\rho)$ destabilizes the interfaces but the term $O(\partial_x^4\rho)$ prevents the gradient from diverging, with a role similar to the positive bending stiffness in the Helfrich Hamiltonian of membranes~\cite{helfrich_elastic_1973,bitbol_bilayer_2012}. This leads to density overshoots and spatial oscillations at the edges of the densely populated domain, but the dense phase ultimately reaches a plateau, as can be observed in snapshots (Figs.~\ref{fig:utilities_snapshots}, \ref{fig:compare_profile_phaseDiag} and \ref{fig:mean-field_local_nonlocal}). Conversely, a strictly positive $\kappa(\rho)$ suppresses the density overshoots, as illustrated in  Fig.~\ref{fig:positive_kappa_divergence}(b). More generally, we believe that such a macroscopic feature, namely, a spatially oscillating density profile could interestingly be exploited in experimental systems to provide important clues on the microscopic properties of the constituents under study.

Coming back to $(\kappa R')'=0$ with $\kappa(\rho^\star)=0$ and assuming that $R(\rho)$ simply needs to be bijective and continuous, one deduces that different constants are now \textit{a priori} needed on each interval $(0,\rhostar)$ and $(\rhostar,1)$ to compute $R(\rho)$. For the specific class of utility function $u(\rho)=-|\rho-\rhostar|^\alpha$, $\alpha \neq 2$, one obtains
\begin{align}
    R(\rho)=\begin{cases}
       \displaystyle\frac{C_1 (\rhostar-\rho)^{2-\alpha}}{\sigma^2 f'_\rational(0)\alpha(\alpha-2)} +C_3\text{ if $\rho<\rhostar$,}\\
   \displaystyle \frac{C_2(\rho-\rhostar)^{2-\alpha}}{\sigma^2 f'_\rational(0)\alpha(\alpha-2)} +C_4\text{ if $\rho>\rhostar$},\label{eq:R_vs_rho_general1}
    \end{cases}
\end{align}
with $C_i$ the interval dependent constants.
These equations show that the case $\alpha<2$ may admit an acceptable change of variable. For $\alpha\geq2$, the function $R$ displays a divergence at $\rho=\rhostar$, which makes impossible to recover an homeomorphism $R:\rho\mapsto R(\rho)$ on the whole domain $(0,1)$. 

In the next paragraph, we detail the procedure introduced in~\cite{solonPRE2018,solon_generalized_2018} to recover the binodal densities. We show that we can extend the procedure to negative $\kappa(\rho)$ -- assuming that higher order gradient terms stabilize the interface. This is one of the central results of this paper.

\begin{figure}
    \includegraphics[width=\linewidth]{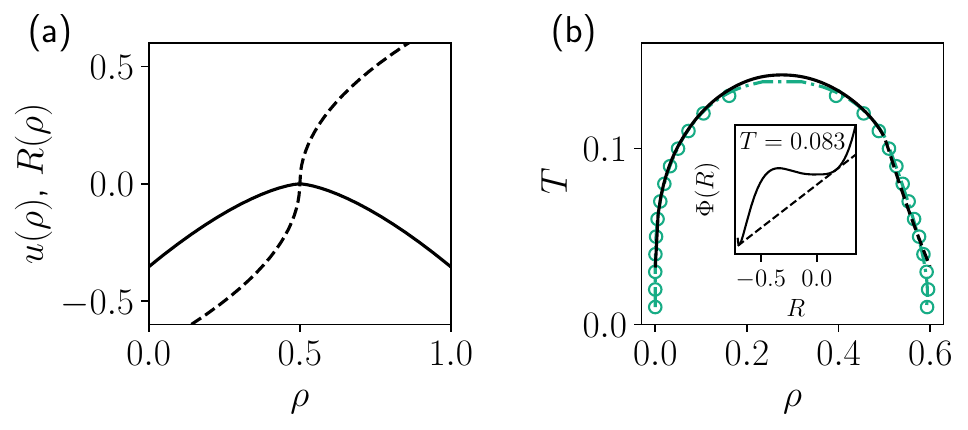}
    \caption{(a) Utility $u(\rho)$ (solid line), and change of variable $R(\rho)$ (dashed line) for $\alpha=3/2$ and $\rho^\star=1/2$. 
    (b) Comparison between the semi-analytical prediction (dark line) and the binodal densities both obtained via the Monte Carlo simulations (green circles) and  solving numerically the mean-field Eq.~\eqref{eq:MF_hydro} (green dot-dashed line). The decision rule here is the logit function, whose values in $0$ are $f_\rational(0)=1/2$ and $f'_\rational(0)=\rational/4$. Inset: double-tangent construction on $\Phi(R)$ for $T=0.083$. The dense phase is given by $R_d=0.170$, yielding $\rho_d=0.53$, and the gaseous phase is given by $R_g=-0.69$, yielding $\rho_g=0.022$.}
\label{fig:gen_thermo_results}
\end{figure}

The function $R$ is bijective, and can thus be inverted to get $\rho(R)$, and this allows us to define a new chemical potential $g[R]\equiv\mu[\rho(R)]$. The functional $\mathcal G[R]$ is then obtained by integrating $g[R]$ on each domain and by gluing together the two integrated parts at $\rhostar$. Explicitly using the notations introduced in~\cite{solon_generalized_2018}, we have
\begin{align}
    g=\frac{\delta \mathcal G  }{\delta R}, \qquad \mathcal G = \int \dd x \left[ \Phi(R) + \frac{\kappa}{2R'}(\partial_xR)^2 \right],
\end{align}
 where $\Phi(R)$ defines a generalized free energy density verifying
\begin{align}
    \frac{\dd\Phi}{\dd R}=g_0[\rho(R)],
\end{align}
with $g_0$ defined in Eq.~\eqref{eq:g0}.
The double-tangent construction on $\Phi(R)$ then yields the binodal densities.
To be more explicit, from Eq.~\eqref{eq:R_vs_rho_general1}, we assume that $R$ is simply given by
\begin{align}
    R(\rho)=\begin{cases}
       -(\rho^\star-\rho)^{2-\alpha}\text{ if $\rho<\rhostar$,}\\
    (\rho-\rho^\star)^{2-\alpha}\text{ if $\rho>\rhostar$},
    \label{eq:R_vs_rho_general2}
    \end{cases}
\end{align}
which can be inverted into $\rho(R)=\rho^\star+\sgn(R)|R|^{\frac{1}{2-\alpha}}$. Injecting $\rho(R)$ in Eq.~\eqref{eq:g0}, one obtains
\begin{align}
\begin{split}
    g_0(R) =&\,2f'_\rational(0)|R|^{\frac{\alpha}{2-\alpha}}\\
    &+f_\rational(0)\log\left[ \frac{\rho^\star+\sgn(R)|R|^{\frac{1}{2-\alpha}} }{1-\rho^\star-\sgn(R)|R|^{\frac{1}{2-\alpha}}}\right].
\end{split}
\end{align}
The explicit formula for $\Phi(R)$ is more involved and we choose to display $R(\rho)$ in Fig.~\ref{fig:gen_thermo_results}(a) for $\alpha=3/2$ and $\rho^\star=1/2$.
To obtain the binodal densities, one either performs the double-tangent construction on $\Phi$ (Fig.~\ref{fig:gen_thermo_results}(b), inset) or the Maxwell construction, the latter being easier to handle numerically~\cite{solon_generalized_2018}. We provide details on these two constructions in Appendix~\ref{app:double_tangent}. In a nutshell, one obtains the coexistence densities of the dilute and the dense phases from the constraints of the steady state. Indeed, in the steady state, the interface between the phases does not move and each phase has a fixed density, which translates into equality across phases of pressure and chemical potential, respectively. We then compare the predictions to the phase densities measured in MC simulations in Fig.~\ref{fig:gen_thermo_results}(b), and report an excellent match.  Note that the interaction range $\sigma$, that appears in $\kappa$ but not in $\Phi$, does not play a role in the predicted coexistence densities of the infinite size system, confirming that the sub-optimal aggregation of agents in a dense cluster is not limited to finite size lattices.

As a final remark, and to provide some validity criterion for the extended change variable, we should mention that $R(\rho)$ taken alone does not contain information on the sign of $\kappa(\rho)$. This is why we claim that, if the system indeed undergoes a true phase separation and that $\nabla\rho$ has finite left and right limits at each point of the domain, then the sign of $\kappa$ does not matter and a computation with negative $\kappa$ still predicts the correct binodal densities. In other words, it is not the negative sign of $\kappa$ \emph{per se} that leads to the failure of the gradient expansion, but rather the fact that there exists some points $x_c$ in the domain such that $|\partial_x \rho\small|_{x_c^{\pm}}|=+\infty$. The fact that $\partial_x \rho$ may not be continuous but keeps finite values at the right and at the left of each point of the domain does not seem to be an issue when performing the gradient expansion, probably because the set of points where the derivative is higher than some arbitrary value is a set of zero measure. The breakdown of the gradient expansion for a positive and monotonic $\kappa(\rho)$ is illustrated in App.~\ref{app:gradient_breakdown}.

\section{Further socioeconomic considerations}
\label{sec:extensions}

\subsection{Two populations}
\label{sec:two_pop}

 A natural extension of the problem is to restore some diversity among agents, as initially considered by both Sakoda and Schelling. Here we consider two types of interacting agents (say $A$ and $B$), with possibly different utility functions, which could for example represent higher and lower revenue individuals, or city dwellers and business professionals, etc.\footnote{The recent work \cite{seara2023sociohydrodynamics} on urban segregation in the United States has brought to our attention the existence of surveys~\cite{farley_continued_1993, clark_ethnic_2002, bobo_attitudes_1996} confirming the idea that different sub-populations may require markedly different utility functions.} A central question in this case is whether the system reaches fixed points, or if more complicated dynamics can persist in the long time limit, especially if the two populations have competing interests. Recent work has been devoted to studying nonreciprocal interactions between different kinds of particles, exhibiting the wealth of possible dynamical behavior when particle displacements are local~\cite{saha_scalar_2020,dinelli_2022}. An interesting question in our setup is for instance: do propagating waves (or frustrated states) survive when nonlocal moves are allowed? Indeed, one may expect that enforcing local displacement constitutes a dynamical constraint that drives the system in a particular way. Allowing for nonlocal moves may change the dynamics of how the frustrated states are resolved.  
  One may think of three major types of interactions:
\begin{itemize}
    \item 
First, a cooperative interaction where agents $A$ and agents $B$ may maximize their utility when agents of opposite type are found in their neighborhood. This kind of interaction will typically lead to homogeneous well-mixed systems, or to some condensation into a dense phase where agents are well-mixed, but since frustration is not implemented in the microscopic rules, we reasonably expect stationary states. 
  \item Second, each agent type may decide to settle among peers and/or avoid agents of the other type in their surroundings. One should then expect a complete phase separation into two domains, one displaying a majority of $A$s and, the other, a majority of $B$s. Whether the $A-B$ phase separation additionally displays some condensation depends on the self-affinity of each agent type.
  \item Third,  frustrated situations in which $A$ settles with $A$ but wants to avoid $B$ agents, while $B$ agents would like to gather and settle close to $A$. In this situation, we may expect non stationary patterns, stemming from the fact that all agents cannot be satisfied at the same time.
\end{itemize}

With this last situation in mind, we have considered the following utility functions ($u_A$ for $A$ agents and $u_B$ for $B$ agents):
\begin{align}
    u_A(x, [\phi_{A,B}]) &=-|\phi_A(x)-\rho^\star|^2 +c_1\phi_B(x)\\
    u_B(x, [\phi_{A,B}]) &=-|\phi_A(x)-\rho^\star|^2 + c_2\phi_B(x),
\end{align}
where $c_1<0$ translates the fact that $A$s are fleeing from $B$, and $c_2>0$ translates the fact that $B$s have a tendency to gather with $B$s. The term $-|\phi_A-\rho^\star|^2$ enjoins both populations to settle among $A$ populated areas. Of course, the specific shape of utilities taken here may be restrictive and can be easily generalized.

The extension of the mean-field dynamics to this two population problem is rather straightforward. Writing $\rho_A(x,t)$ (resp. $\rho_B(x,t)$) the density of agents $A$ (resp. $B$) at location $x$ and time $t$, and denoting the total density by $\rho(x,t)\equiv \rho_A(x,t)+\rho_B(x,t)$, we now have an evolution equation of the form
\begin{equation}
\begin{aligned}
    \partial_t\rho_A(x,t) &= [1-\rho(x,t)] \int \rho_A(y,t)w_{\rational_A}([\phi_{A,B}],y,x,t) \, \dd y\\
    &- \rho_A(x,t) \int [1-\rho(y,t)] w_{\rational_A}([\phi_{A,B}],x,y,t) \, \dd y,
\end{aligned}
\label{eq:MF_hydro_2pop}
\end{equation}
and, by symmetry, a similar equation for $B$. The transition rates depend on the utility function of each agent type and are \textit{a priori} agent specific. Denoting $u_Z(x)\equiv u_Z(x,[\phi_{A,B}])$ (with $Z=A$  or $B$), we set
\begin{align}
        w_{\rational_Z}([\phi_{A,B}],y,x,t)= \omega_Z f_{\rational_Z}[u_Z(x)-u_Z(y) ],
\end{align}
where $\omega_Z$ and $\rational_Z$ can be agent dependent.
\begin{figure}
    \centering
    \includegraphics[width=\linewidth]{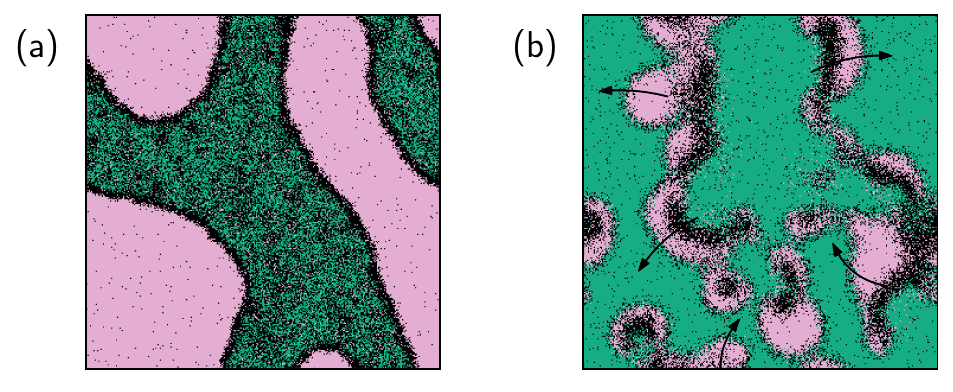}
    \caption{Snapshots of the system for two frustrated interaction parameter choices. (a) Stationary demixing in a region where LSA presents complex eigenvalues. The agent $A$ phase still contains some $B$ agents. Parameters: $c_1=-2$, $c_2=1$, $\sigma=3$, $\bar\rho_A=0.2$, $\bar \rho_B=0.5$, $\rational=10$. (b) Chaotic propagation of polarized blobs in a region where LSA presents pure real eigenvalues (null imaginary part). Parameters: $c_1=-2$, $c_2=0.5$, $\sigma=7$, $\bar\rho_A=0.6$, $\bar\rho_B=0.2$, $\rational=100$. For both (a) and (b), $L_x=L_y=300$. Movies are available online \cite{movies_SM}.}
    \label{fig:demixing_chasing}
\end{figure}

In App.~\ref{app:LSA_two_populations}, we perform the linear stability analysis of the homogeneous state.  As expected, in the frustrated two-population system, unstable modes can display temporal oscillations. However, these oscillations may stop when nonlinear terms become relevant, and the system may end up in a stationary phase separation (similar to classical demixing in equilibrium systems), as displayed in Fig.~\ref{fig:demixing_chasing}(a). 
Reciprocally, non-oscillating growing modes at the linear level may give rise to propagating structures and waves when nonlinearities become important, as shown in Fig.~\ref{fig:demixing_chasing}(b) (see Supplementary Material  in~\cite{movies_SM}). In our system, and at odds with recent work~\cite{saha_scalar_2020,dinelli_2022}, the oscillatory nature of the non-homogeneous steady state cannot be predicted from a simple linear stability analysis about the homogeneous solution.

A thorough analysis of the emerging behaviors in the multi-population system would require more work, beyond the scope of the present paper. Still, it is remarkable that, here as well, the linear stability analysis in the case of local jumps yields exactly the same instability conditions as the nonlocal dynamics ones (see results of Appendices \ref{app:LSA_two_populations} and \ref{app:LSA_two_populations_local}). As a consequence, we expect that some results of the recent works~\cite{saha_scalar_2020,dinelli_2022} should be relevant, to some extent, to describe our multi-population system.

\subsection{Housing market}
\label{subsec:housing_market}

A common and reasonable criticism of the kind of model developed here is that, while the perceived density may be a significant factor in the decision making process of agents, the price of a house should also necessarily be taken into account. Indeed, in classical economics, the market is usually considered to be the mechanism through which the optimal allocation of goods and assets occurs (despite some contradicting empirical evidence e.g. \cite{gabaix2021search}). As a result, one could rightfully argue that a housing market is necessary to ensure that agents eventually reach a steady state where their utility is maximal, at odds with what we have observed in the condensed phase.

Incorporating pricing in the model is not trivial, however, and there are a number of ways in which this could be done. A common approach in the modeling of socioeconomic systems is to introduce an agent-dependent budget and to constrain agents' moves based on such budget, as done in \cite{gauvin_modeling_2013} for example. Realistically, this budget should then be heterogeneous among the agents (e.g. Pareto distributed for instance). While relevant and interesting, this agent-specific dependence as well as its formulation as a hard constraint would require a different modeling approach, and is unlikely to be tractable analytically. The alternative that we take here is to consider that if a given move leads to excess costs for an agent, its utility would decrease.
 We may then conveniently stay within the modeling framework of our model and assume that such a price field $\psi$ is an increasing function of the smoothed density field $\phi$, such that houses are more expensive in dense neighborhoods-- or in other words: prices are  driven by demand only. In its most general form, we propose the price-adjusted utility
\begin{equation}
     \bar{u}(\phi)= u(\phi) - u_\mathrm{p} [\psi(\phi)],
     \label{eq:utility_tilde_price}
\end{equation}
where $u_\mathrm{p}$ is the price penalty, assumed to be an increasing function of the price, and therefore, of the density $\phi$. This penalty is then, by construction, expected to drive the system away from condensation.  

For concreteness, one can consider the example of a linear penalty term on the utility function , and  that the price grows linearly with the local (smoothed) density, such that:
\begin{equation}
     \bar{u}(\phi)= u(\phi) - \gamma \phi,
\end{equation}
with $\gamma>0$.
Interestingly, introducing such a coupling boils down to introducing a pair-wise repulsive interaction between agents. The condition to observe condensation is then shifted by $\gamma$, and reads explicitly
\begin{equation}
     2 \rho_0(1-\rho_0)\frac{f'_\rational(0)}{f_\rational(0)}(u'(\rho_0)-\gamma)> 1.
     \label{eq : new_criterion}
\end{equation}
Clearly, condensation cannot occur anymore if the price penalty on the utility is too important.

This example therefore illustrates that it is indeed possible for an appropriate housing market to destroy the condensation observed in our continuous model. However, it is important to note that the outcome (without the price penalty) for agents is not necessarily improved by this brutal homogenization through the price field. Besides, it can be argued that the effect of price should not be as trivial as a utility penalty. In fact, other models have used price as a proxy for social status, in which case agents are actually more attracted by the most expensive site they can afford \cite{gauvin_modeling_2013}. In App.~\ref{app:complex_price_field} we also consider the case of a more complex price field dependence. Ongoing research is devoted to performing a comprehensive study of housing price correlations in empirical data and the careful construction of an adequate pricing dynamics~\cite{becharat2023}.

\section{Discussion}
\label{sec:discussion}
Let us summarize what we have achieved in this paper. We have introduced a neighborhood-less extension of the Sakoda-Schelling model for the occupation of a lattice representing a city. In this version of the model, the agents attempt to maximize a utility that is a function of their \textit{perceived} local density, and are most satisfied when they are located in a site where such density is of an intermediate value, i.e. neither empty nor too crowded. Having that  agents only consider their own site dependent utility, and that their utility is nonlinear, drives the system out of equilibrium. As a result, the system can no longer be studied by constructing a free energy directly from an aggregate system-wide utility function, as was done by Grauwin \textit{et al.}~\cite{grauwin_bertin_2009} for agents inhabiting predefined neighborhoods or blocks in which the utility is identical for all. Using numerical simulations as well as a mean-field description of the nonequilibrium dynamics, we have established that the apparent disparity between \textit{micromotives} and \textit{macrobehaviours} initially observed by Schelling \cite{schelling1971} is robust to the absence of neighborhoods and to the out-of-equilibrium nature of our extension. Interestingly, we find that the transition between the fully homogeneous state and the phase-separated one likely belongs to the 2D Ising universality class, a debated topic in the active matter literature regarding the very similar Motility Induced Phase Separation (MIPS) phenomenon. Taking advantage of the similarity between our problem and the Active Model B (describing MIPS), we predict the local density in the bulk of the concentrated phase, confirming that the majority of agents find themselves in a sub-optimal situations with a perceived density exceeding the ideal value. 

While seemingly technical, the fact that the original observations of Schelling is robust to out-of-equilibrium dynamics actually carries far reaching consequences, in our opinion. Indeed, as discussed in Sec.~\ref{sec:out_of_equ}, equilibrium descriptions of socioeconomic problems require the decision rule $f_\rational(x)$ to be the ``logit'' function. This very specific choice is a common source of criticism, as any results are then \textit{a priori} uncertain to hold for other decision rules. Here, on the other hand, our out-of-equilibrium description presents no such restriction, as all calculations have been written as generally as possible and, interestingly, turn out to only depend on $f_\rational(0)$ and $f'_\rational(0)$. While the numerical results presented here are those using the classical choice of the logit function for lack of a more plausible decision rule, one could readily adapt the outcomes following behavioral evidence that another function is more appropriate, and we expect the entire phenomenology of our model to hold for any other sigmoid function. More generally, we believe that this robustness to other decision rules holds for a large number of socioeconomic models that have been described using the methods of statistical physics \cite{garnier2022bounded,Blume1993,brock1993pathways,borghesi2007songs,moran2020force}. Of course, subtleties around the dynamics, such as the relaxation time towards the steady state or the coarsening dynamics discussed here, will inherently be affected by the specific choice that governs transition rates. This being said, we have observed a remarkable similarity in the local and non-local versions of our model for which the dynamics are yet qualitatively very different. It is therefore possible that there also exists some degree of universality in the dynamical behavior of different decision rules, at least at the mean-field level.

Going back to the Sakoda-Schelling model, we have also introduced some generalizations that we believe are natural and relevant. First, the introduction of different sub-populations is interesting, as the exhibited dynamical patterns are impossible to observe in an equilibrium version of the model.
Second, we have seen that introducing a linear dependence of the price on the density has the effect of delaying the transition, eventually killing it off completely if the price penalty in the utility function is strong enough. As previously stated, however, this mechanism remains very simple. In order to determine more plausible effects of a housing market, a thorough analysis of real estate transactions appears to be a promising avenue, in particular given the increasing availability of open datasets in this area in major European cities. An extensive study of French data is currently underway, hopefully allowing us to couple this continuous Sakoda-Schelling model with a plausible housing market model in the near future~\cite{becharat2023}. Note finally that the recent preprint~\cite{seara2023sociohydrodynamics} revisits the problem of urban segregation in the United States and proposes a two-population model very similar to ours. Studying the validity of hydrodynamic descriptions of the two-population problem using the census data brought forth in this paper appears to be an important perspective. Such a comparison could notably be necessary to highlight the role of ingredients not taken into account in existing models -- such as public policy and economic inequalities -- in the emergence of strong geographic disparities.

\section*{Acknowledgements}
We warmly thank Jean-Philippe Bouchaud for his numerous insights on this study, as well as Claire Alais and No\'e Beserman who participated in the early stages of this project. We also thank Eric Vanden-Eijnden for fruitful discussions, as well as Eric Bertin for useful comments on the manuscript. R. Z. also thanks J\'er\'emy O'Byrne for precious discussions along the years on thermodynamic mappings. J. G.-B. finally thanks Samy Lakhal for fruitful discussions on the linear utility case. This research was conducted within the Econophysics \& Complex Systems Research Chair, under the aegis of the Fondation du Risque, the Fondation de l’\'Ecole polytechnique, the \'Ecole polytechnique and Capital Fund Management.

\appendix

\section{Coarsening exponent}
\label{app:coarsening}

We start from a homogeneous system of size $L_x\times L_y$ with $L_x=L_y$, and we quench it below the critical temperature in the spinodal region.
The system undergoes a spinodal decomposition where dense domains coarsen until forming one single large cluster. The typical size of the domains, denoted $L_d$ grows with time as $\sim t^{1/z}$, where the growth exponent $z$ indicates the physics at play. To measure the typical domain size, we compute first the structure factor, given by
\begin{align}
    S(\bm k, t) = \Big|\sum_{\bm r} e^{-i\bm k\cdot \bm r} \phi(\bm r,t)\Big|^2.
\end{align}
Using isotropy of the system, we average the structure factor over given shells $q=(k_x^2+k_y^2)^{\frac 12}$ and we obtain the radial structure factor $s(q,t)=\int_{[0,2\pi]}S(q,\theta,t) \dd\theta $. The typical domain size is given by
\begin{align}
    L_d(t) = 2\pi\dfrac{  \int_{k_1}^{\Lambda} s(q,t) \dd q }{ \int_{k_1}^{\Lambda} q \,s(q,t) \dd q},
\end{align}
with $\Lambda$ the ultraviolet cutoff and $k_1=2\pi/L_x$ the infrared cutoff. On our finite grid, the integral takes the form of a discrete sum, the wavenumber $q$ ranges from $2\pi/N_x$ to $2\pi(N_x-1)/N_x$, and the increment $\dd q$ is replaced by $2\pi/N_x$. 

\begin{figure}[b]
    \includegraphics[width=0.8\linewidth]{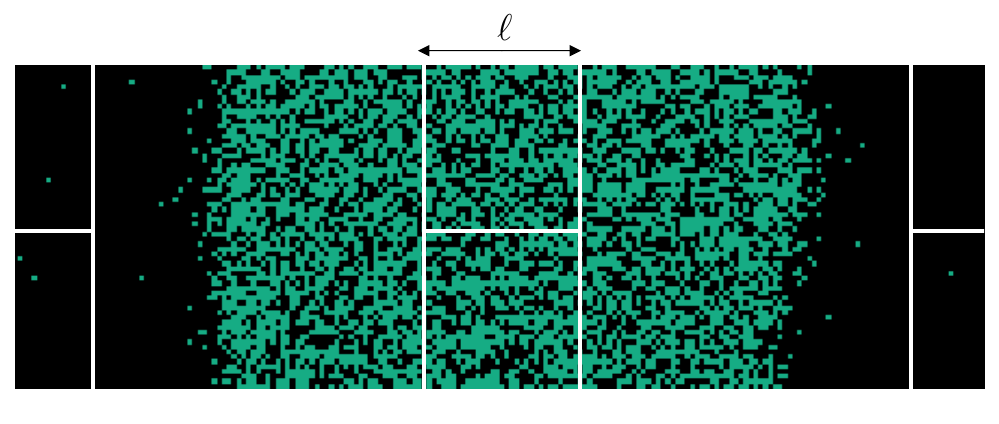}
    \caption{Snapshot of a Monte Carlo simulation. Green sites are occupied, black sites are empty. We draw the boxes of size $\ell=L_x/6$ that are used to measure the liquid and the gas densities. Here, the system size is $L_x=200$, $L_y=66$. Parameters: $\alpha=3/2$, $\rho_0=0.35$, $T=0.05$.}
    \label{fig:boxes_centered}
\end{figure}

\begin{figure}
    \includegraphics[width=\linewidth]{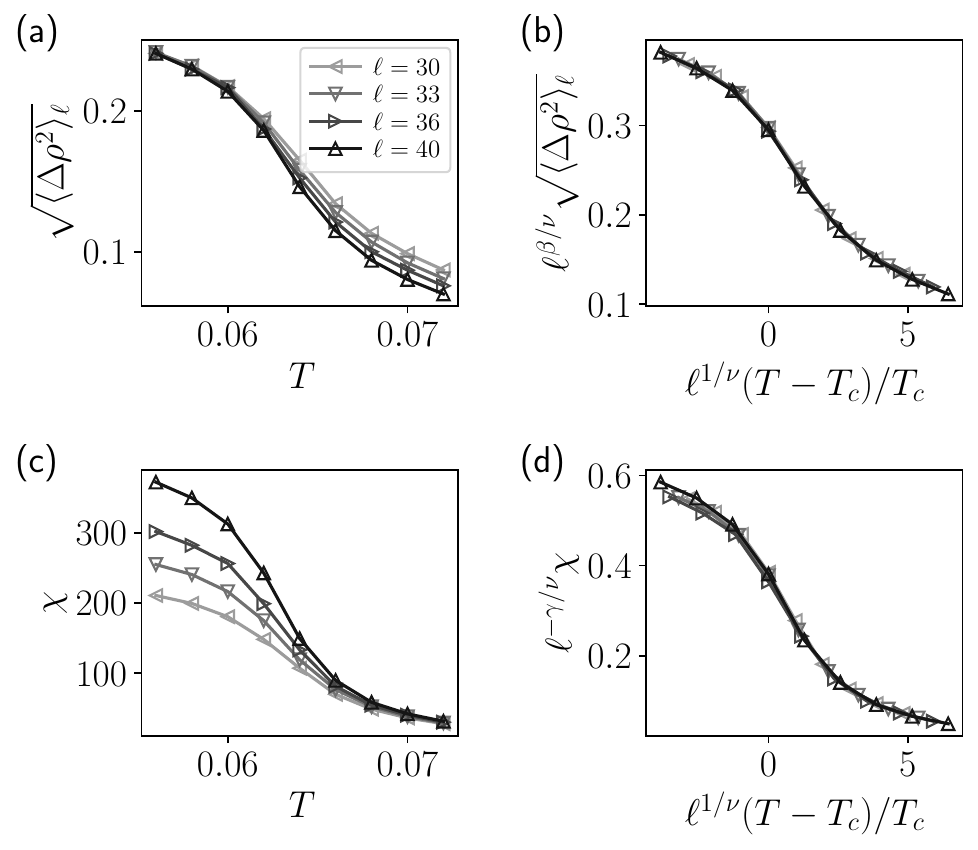}
    \caption{Binder cumulant, order parameter and compressibility close to the critical point $(\rho_c, T_c)=(0.271,0.0620)$ computed for $\alpha=3/2$ and $\sigma=1$, as a function of the temperature $T$.}
    \label{fig:Binder_collapse}
\end{figure}

\section{Monte Carlo simulations for 2nd-order phase transitions}
\label{app:critical_exponents_add}

The Monte Carlo simulation setup that we used is shown in Fig.~\ref{fig:boxes_centered}. As introduced in recent works on MIPS, we study the phase transition using four boxes of size $\ell \times \ell$ located in the bulks of the dense and ``gas'' phases. The initial condition for the simulations is a fully separated state, where a slab of density 1 coexists with a slab of density 0. We measure the system decorrelation time $\tau_d$ and we start recording data after $\sim1.5\tau_d$. Each simulation is run for a time $\geq 5\tau_d$, and each symbol in Fig.~\ref{fig:Binder_collapse} aggregates the data of $80$ independent simulations.

The collapse of the different observable with the 2D Ising critical exponents is displayed in Fig.~\ref{fig:Binder_collapse}.

\section{Lyapunov function for non-local moves}
\label{app:lyapunov_nonlocal}

We show below that, in the mean-field limit, with the logit decision rule and a linear utility function, the hydrodynamic evolution has a Lyapunov function.
Starting from the pairwise Hamiltonian 
\begin{equation}
\mathcal{H} = -\frac{\nu}{2} \sum_{\bm r, \bm r'} n(\bm r) G_\sigma (\bm r- \bm r') n(\bm r'),
\end{equation}
which can be shown to satisfy Eq.~\eqref{eq:linkage} when $u(\phi) = \nu \phi$, we take the continuous-space limit and we account for the entropic contribution to find the free energy functional
\begin{eqnarray}
\mathcal{F}[\rho] = -\frac{\nu}{2} \int \dd x \, \dd y  \, \rho(x) G_\sigma(x-y) \rho(y)  +  T S[\rho],\label{eq:free_energy_linear}
\end{eqnarray}
with the entropy $S[\rho] = \int [\rho \log(\rho) + (1-\rho)\log(1-\rho)]$.
Before addressing the dynamics of non-local moves, let us address the one of local moves. It turns out that when moves are local, the use of a linear utility and the logit rule implies that the mean-field dynamics is of gradient type, analogous to a Wasserstein gradient flow but with $M[\rho]=\rho(1-\rho)$ instead of $M[\rho]=\rho$~\cite{otto_geometry_2001}, with $\mathcal{F}$ playing the role of the free energy potential. Naturally, the stochastic dynamics is in detailed-balance with the Gibbs-Boltzmann measure $e^{-\rational \mathcal F}$. When the moves are no longer local, the gradient structure at the mean-field level is difficult to unveil, even though the stochastic dynamics remains in detailed-balance. The functional $\mathcal F$ remains a Lyapunov function of the dynamics, i.e. it is non-increasing as the dynamics evolves. This is what we show explicitly in the next paragraph.

For simplicity, we define the chemical potential \begin{eqnarray}
\mu(x)\equiv\frac{\delta \mathcal F}{\delta \rho(x)}=-\nu \phi(x)+T\log\Big(\frac{\rho(x)}{1-\rho(x)}\Big),
\end{eqnarray}
such that inverting the equation yields
\begin{align}
\rho(x)=\frac{e^{\frac{\rational}{2} (\mu(x)+\nu\phi(x)) }}{D(x)},
\label{eq:rho_mu_lyapunov}
\end{align}
with $D(x)\equiv 2\cosh[\frac{\rational}{2}(\mu(x)+\nu\phi(x))]$.
Then, computing the total time derivative on the functional yields (we omit the time dependence of the fields to alleviate notations):
\begin{align}
\begin{split}
    \frac{d\mathcal F}{dt}=&\int_x  \frac{\delta \mathcal F}{\delta \rho(x)}\partial_t \rho(x,t)\dd x\\
    =& \int_x  \mu(x) (1-\rho(x,t)) \int_y \rho(y,t)w_\rational([\phi],y,x,t) \dd y \dd x \\
    &- \int_x\mu(x)\rho(x,t) \int_y (1-\rho(y,t)) w_\rational([\phi],x,y,t) \dd y\dd x\\
=& -\iint \dd  x \dd y\,  \frac{ Z(x,y)}{4D(x)D(y)\cosh[\frac{\rational}{2}(\phi(x)-\phi(y))]},
\end{split}  
\label{eq:df_dt_Lyapunov}
\end{align}
where we have replaced $w_\rational$ by the logit decision rate [see Eq.~\eqref{eq:logit_rate}] and we have symmetrized and simplified the second line to obtain the third line, and where we define $Z(x,y)\equiv[\mu(y)-\mu(x)](e^{\frac{\rational}{2}(\mu(y)-\mu(x))}-e^{\frac{\rational}{2}(\mu(x)-\mu(y))})\geq 0$, $\forall x,y$.
 We thus conclude that $\frac{d\mathcal F}{dt}\leq 0$, i.e. that $\mathcal F$ is a Lyapunov function of the hydrodynamic evolution when the utility is linear.
Now, more than that, the function is always strictly decreasing unless it starts from a fixed point, which forbids limit cycles. Indeed, one notices that the integrand in Eq.~\eqref{eq:df_dt_Lyapunov} is always positive, unless $\mu(x,t)=\mu(y,t)$, for all $x,y$. Setting $C(t)=\mu(x,t)$ on the manifold where $\mathcal F$ is constant, we have
\begin{align}
    1-\rho(x,t)=\rho(x,t)e^{\Gamma C(t)}e^{\Gamma\nu\phi(x,t)}.
\end{align}
If we inject this relation in the non-local mean-field evolution Eq.~\eqref{eq:MF_hydro}, then we obtain $\partial_t\rho(x,t)=0$, indicating that $\rho(x,t)$ must be a stationary fixed point when the Lyapunov function is constant.

\section{Local mean field description and LSA}
\label{app:LSA_localMoves}

In this section, we consider a modified dynamics where agents are allowed to relocate on neighboring site only. For simplicity, we also consider that the system is one dimensional.
It is thus possible to perform a Taylor expansion of the different fields assuming that all fields are smooth in the mean-field limit. The jump probability between two neighboring sites becomes
\begin{align}
    f_\rational[u(x+a)-u(x)] = f_\rational\left(a \partial_x u +\frac{a^2}{2}\partial_x^2u\right),
\end{align}
where $a$ is the lattice size, and $u$ is the utility on position $x$.
The evolution of the density (for non-overlapping agents) is thus given by
\begin{align}
\begin{split}
    \partial_t \rho =& \rho(x+ a )[1-\rho(x)]f_\rational(-a\partial_x u-\frac{a^2}{2}\partial_x^2 u)\\
    &+\rho(x-a )[1-\rho(x)]f_\rational(a\partial_x u-\frac{a^2}{2}\partial_x^2 u)\\
    &-\rho(x)[1-\rho(x+a )]f_\rational(a\partial_x u+\frac{a^2}{2}\partial_x^2 u)\\
    &-\rho(x)[1-\rho(x-a )] 
    f_\rational(-a\partial_x u +\frac{a^2}{2}\partial_x^2 u).
    \end{split}
\end{align}
After Taylor expansion up to $O(a^2)$ and time rescaling, it turns out that the evolution equation simplifies into 
\begin{align}
    \label{equ:mean_field_local}
    \partial_t \rho &= f_\rational(0) \partial_x^2 \rho -2f_\rational'(0) \partial_x[\rho(1-\rho)\partial_x u].
\end{align}
Then, expanding around an homogeneous state, we write $\rho=\rho_0+\rho_1(x,t)$, $\phi=\rho_0+\phi_1(x,t)$, and we obtain to leading order in the perturbation:
\begin{equation}
    \partial_t \rho_1=f_\rational(0)\partial_x^2\rho_1-2f_\rational'(0)\rho_0(1-\rho_0)u'(\rho_0)\partial_x^2\phi_1.
\end{equation}
In Fourier space the evolution of the mode $k$ is given by $\partial_t\hat \rho_1=\Lambda(k)\hat \rho_1$, with 
\begin{equation}
     \Lambda(k)=-k^2f_\rational(0)\left(1 
     -2\frac{f'_\rational(0)}{f_\rational(0)}\rho_0(1-\rho_0)u'(\rho_0)\hat G_\sigma(k)\right).
\end{equation}
From this, we deduce that the homogeneous system is unstable if there exists a mode $k^\star$ such that 
\begin{align}
    1<2\frac{f'_\rational(0)}{f_\rational(0)}\rho_0(1-\rho_0)u'(\rho_0)\hat G_\sigma(k^\star).
    \label{eq:lsa_local_criterion}
\end{align}
This criterion is exactly the same as the one found for the non-local move dynamics.

\section{Local versus non-local PDEs}
\label{app:local_v_nonlocal}
\begin{figure}[h!]
\includegraphics[width=\linewidth]{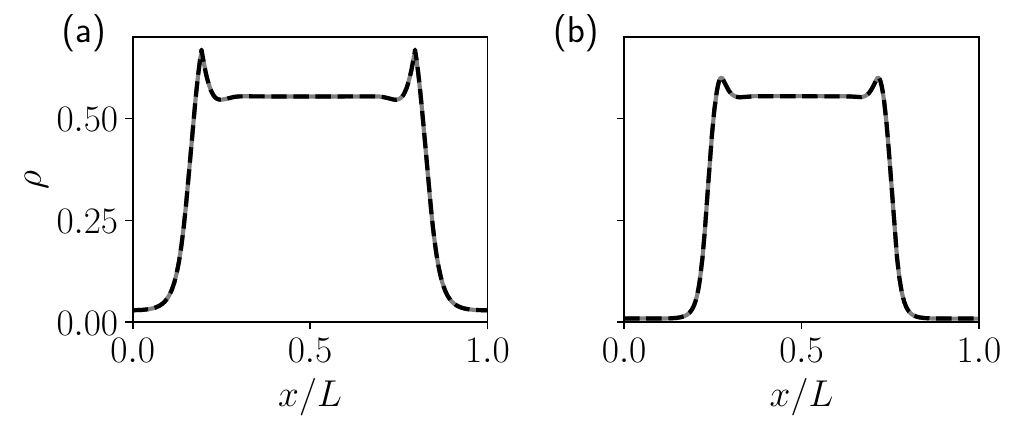}
  \caption{1D steady-state profiles of the density $\rho$ computed by solving the mean-field dynamics with local moves (solid line), or non-local moves (dashed line). The two density profiles superimpose almost exactly, independently of the various parameter values. It was checked that different initial configurations lead to the same final state.
  a) Parameters: utility exponent $\alpha=1$, $\rho_0=0.4$, $\sigma=12$, $L=300$, $\rational=9$.  b) $\alpha=3/2$, $\rho_0=0.3$, $\sigma=10$, $L=300$, $\rational=15$.}
  \label{fig:mean-field_local_nonlocal}
\end{figure}

To illustrate the effectiveness of our local-move approximation in the description of the steady state of the system, we have solved numerically both the local and the non-local mean-field PDEs for the same parameters. The resulting density profiles, displayed in Fig.~\ref{fig:mean-field_local_nonlocal}, appear to be strictly identical up to numerical errors.

\section{Breakdown of the gradient expansion}
\label{app:gradient_breakdown}
\begin{figure}
    \includegraphics[width=\linewidth]{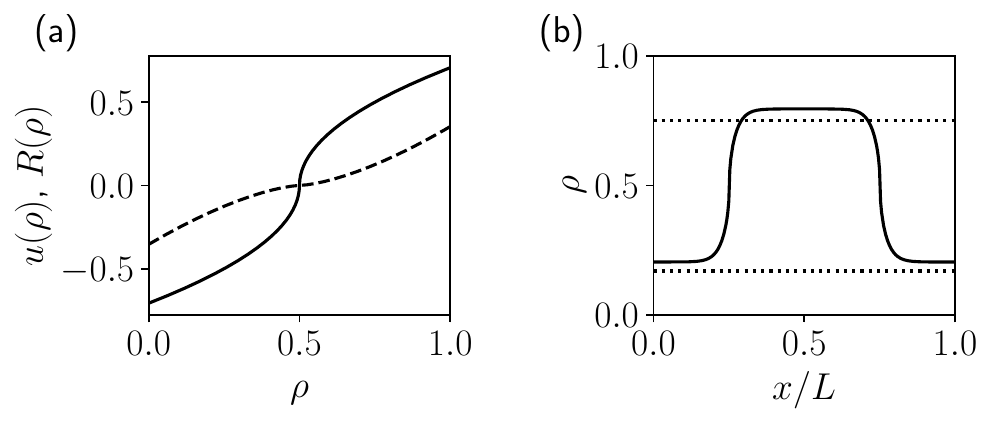}
    \caption{a) Monotonic utility $u(\rho)=\sgn(\rho-\rho^\star)|\rho-\rho^\star|^{1/2}$ (solid line), and bijective change of variable $R(\rho)=\sgn(\rho-\rho^\star)|\rho-\rho^\star|^{3/2}$ (dashed line) for $\rho^\star=1/2$. b) Density profile when phase separation occurs, for $\rational=2.5$, $\sigma=10$, $L_x=300$. The plateaus of the liquid and the gas phase do not match the plateaus predicted by the generalized thermodynamic mapping because the gradient expansion is no longer valid close to $\rho=0.5$. }
    \label{fig:positive_kappa_divergence}
\end{figure}

To illustrate the possible breakdown of the gradient expansion even for $\kappa(\rho) > 0$, we consider a strictly increasing and continuous utility with a $|\rho-\rho^\star|^{-1/2}$ divergence. As shown in the comparison with the mean-field PDE solved numerically in Fig.~\ref{fig:positive_kappa_divergence}, the generalized thermodynamics fails at predicting the bulk densities in this pathological case.

\section{Details on the double-tangent and the Maxwell constructions}
\label{app:double_tangent}

For completeness, we provide a summary of the two approaches that yields the binodal densities, as they are presented in~\cite{solon_generalized_2018}.
As stated the main text, when a phase separation occurs in equilibrium, the density profile in the stationary state is no longer evolving, i.e. the free energy, constrained by the fact that the total mass of the field is fixed, has reached a minimum. Since interfaces have a sub-extensive contribution in the thermodynamics limit, one can work with the free energy density $f(\rho)$. The chemical potential it thus $\mu(\rho)=\frac{\dd f}{\dd \rho}$.
In the steady state, one has equality of the chemical potential in the liquid and in the gas, i.e
\begin{align}
    \mu(\rho_\ell)=\mu(\rho_g)=\bar \mu,
\end{align}
where $\rho_\ell$ and $\rho_g$ denote the densities in the liquid and in the gas, respectively.
Since the interface does not move, one also has equality of pressure across the interface, i.e
\begin{align}
    P(\rho_\ell)=P(\rho_g)=\bar P,
\end{align}
with $P(\rho)=\rho \mu(\rho)-f(\rho)$.
We are thus looking for the two densities such that the tangent to the free energy density $f$ in $\rho_\ell$ and $\rho_g$ is the same, with the slope given by
\begin{align}
    \bar \mu = \frac{f(\rho_\ell)-f(\rho_g)}{\rho_\ell-\rho_g}.
\end{align}
Equivalently, the coexisting densities can be obtained via the Maxwell equal-area construction that imposes
\begin{align}
    \int_{v_\ell}^{v_g}[P(v)-\bar P]\dd v=0,
\end{align}
where $v\equiv 1/\rho$ is the volume per particle, and $v_{g/\ell}=1/\rho_{g/\ell}$.
\begin{figure}
    \centering
    \includegraphics[width=0.99\columnwidth]{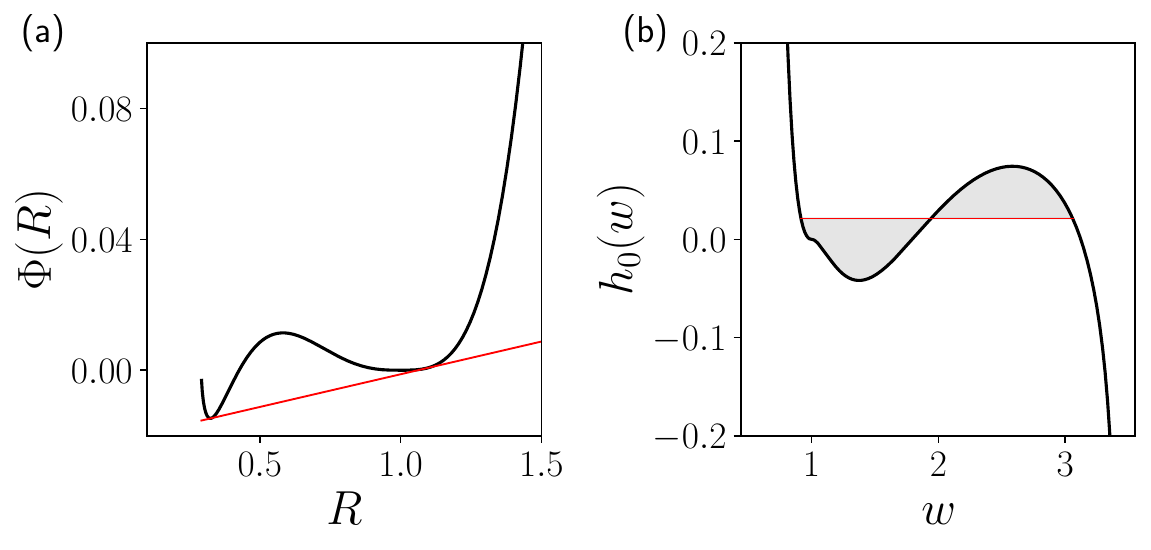}
    \caption{(a) Double-tangent construction (in red) on the function $\Phi(R)$ and (b) Maxwell equal-area construction on $h_0(w)$, with $h_0(w)=\bar h$ in red. Shaded areas in (b) are equal. Here we have plotted these functions for $\alpha=3/2$, $\rho^\star=1/2$, $f_\Gamma(0)=1/2$, $f'_\Gamma(0)=\Gamma/4$, and $\Gamma=10$. We find $\bar h=0.0212$,  $R_\ell=1.092$ and $R_g=0.327$, (or $w_\ell=0.915$ and $w_g=3.058$), translating into $\rho_\ell=0.508$ and $\rho_g=0.047$. }
    \label{fig:tangent_maxwell}
\end{figure}

Here, even though we lie out of equilibrium, we have shown how to find a function $\Phi(R)$ that plays a role similar to a free energy density by means of the change of variable $\rho(R)$. This function is always convex for $T$ above the mean-field critical temperature $T_c^{\mathrm{MF}}$ and display a non-convex region for $T<T_c^{\mathrm{MF}}$.
The function $g_0(R)=\frac{\dd\Phi}{\dd R}$ is now the chemical potential, and the double construction on $\Phi$ imposes 
\begin{align}
g_0(R_\ell)=g_0(R_g)=\frac{\Phi(R_\ell)-\Phi(R_g)}{R_\ell-R_g}.
\end{align}
The equal-area Maxwell construction involves the so-called generalized pressure $h_0$: 
\begin{align}
    h_0(R)=R\frac{\dd \phi(R)}{\dd R}-\Phi(R),
\end{align}
and setting $w=1/R$, we find $w_\ell$ and $w_g$ such that 
\begin{align}
    \int_{w_\ell}^{w_g}[h_0(w)-\bar h] \dd w =0.
    \label{eq:maxwell_condition}
\end{align}
In practice, the function $h_0$ can be obtained numerically and the volume $w_\ell$ and $w_g$ can be obtained with a numerical solver using Eq.~\eqref{eq:maxwell_condition}, more easily than solving the double-tangent condition. We show both constructions in Fig.~\ref{fig:tangent_maxwell}.
Note again that $R$ has no clear physical meaning and is an intermediary variable of computations. As such, one can always choose integration constants in Eq.~\eqref{eq:R_vs_rho_general1} such that $R(\rho)\geq R(0)>0$ to ensure that $w=1/R$ is correctly defined. In the main text though, we have chosen $R(\rho)$ centered in $0$ for simplicity, and computed $g_0(R)$ and $\Phi(R)$ accordingly. If we enforce below $R(0)=1-\rho^\star{}^{2-\alpha}$, then taking $\alpha=3/2$, $\rho^\star=1/2$, $f_\Gamma(0)=1/2$ and $f'_\Gamma(0)=\Gamma/4$, we can obtain an analytic expression for $\Phi(R)$
\begin{widetext}
\begin{align}
\begin{split}
   \Phi(R)=&  \frac{\sqrt{2}}{2} \left(2\Theta(R-1)-1\right) \left[ \frac{R-1}{\sqrt{2}} \log\left(\frac{1+2(R-1)^2}{1-2(R-1)^2} \right) 
   + \arctan\left((R-1)\sqrt{2}\right)
   -\arctanh\left((R-1)\sqrt{2}\right)\right]\\
   &+\frac{\Gamma}{8}   |R-1| (R-1)^3.
   \end{split}
\end{align}
\end{widetext}

\begin{widetext}
\section{Linear stability for two coupled populations}
\label{app:LSA_two_populations}

We consider the evolution of a perturbation of the homogeneous state in Eq.~\eqref{eq:MF_hydro_2pop} (and in its coupled analogue for the field $\rho_B$).
Close to the homogeneous state $\rho_A(x)\equiv\bar \rho_A$, $\rho_B(x)\equiv\bar \rho_B$, with $\rho_0=\bar \rho_A+\bar \rho_B$, we expand the fields $\rho_Z(x,t)=\bar \rho_Z +\rho_{Z,1}(x,t)$, with $Z=A$ or $B$, and the perturbation fields are denoted with index $1$. One also has $\rho(x,t)=\rho_0+\rho_1(x,t)$, and $\phi_Z(x,t)=\bar \rho_Z+\phi_{Z,1}(x,t)$. Keeping leading order terms in Eq.~\eqref{eq:MF_hydro_2pop} yields 

\begin{align}
    \partial_t\rho_{A,1}=\Omega \omega_A \left[ - \bar \rho_A  f_{\rational_A}(0) \rho_1(x,t)+ 2(1-\rho_0) \bar \rho_A f'_{\rational_A}(0)[\phi_{A,1}(x,t)\partial_1 u_A + \phi_{B,1}(x,t)\partial_2 u_A]
    -(1-\rho_0)f_{\rational_A}(0)\rho_{A,1}(x,t)\right],
\end{align}
where $\partial_1 u_A$ is a shorthand notation for $\frac{\partial u_A}{\partial \bar \rho_A}[\bar \rho_A,\bar \rho_B]$. Taking the logistic function $f_{\rational_A}(0)=\frac{1}{2}$, $f'_{\rational_A}(0)=\frac{\rational_A}{4}$, the linear evolution simplifies into
\begin{align}
    \partial_t\rho_{A,1}(x,t)=\frac{\Omega \omega_A}{2} \left[ -\bar \rho_A\rho_1(x,t)   + (1-\rho_0) \bar \rho_A \rational_A [\phi_{A,1}(x,t)\partial_1 u_A + \phi_{B,1}(x,t)\partial_2 u_A]
    -(1-\rho_0)\rho_{A,1}(x,t)\right].
\end{align}
Similarly, we obtain for the evolution of $B$:
\begin{align}
    \partial_t\rho_{B,1}(x,t)=\frac{\Omega \omega_B}{2} \left[ -\bar \rho_B\rho_1(x,t)   + (1-\rho_0) \bar \rho_B \rational_B [\phi_{A,1}(x,t)\partial_1 u_B + \phi_{B,1}(x,t)\partial_2 u_B]
    -(1-\rho_0)\rho_{B,1}(x,t)\right].
\end{align}
Denoting $\hat \rho_Z(k,t)$ the Fourier transform of $\rho_{Z,1}(x,t)$, the evolution equation can be cast in Fourier space into
\begin{align}
    \partial_t \begin{pmatrix}
\hat \rho_A(k,t) \\
\hat \rho_B(k,t) 
\end{pmatrix}=L\begin{pmatrix}
\hat \rho_A(k,t) \\
\hat \rho_B(k,t) 
\end{pmatrix},
\end{align}
with 
\begin{align}
    L=\frac{\Omega}{2}\begin{pmatrix}
     \omega_A(\bar \rho_B-1 +(1-\rho_0)\bar\rho_A\rational_A\hat G_\sigma(k) \partial_1 u_A)  & \omega_A(-\bar\rho_A + (1-\rho_0)\bar\rho_A\rational_A\hat G_\sigma(k) \partial_2 u_A)\\
      \omega_B(-\bar \rho_B +(1-\rho_0)\bar \rho_B\rational_B \hat G_\sigma(k)\partial_1 u_B)  & \omega_B(\bar \rho_A-1 +(1-\rho_0)\bar\rho_B\rational_B\hat G_\sigma(k) \partial_2 u_B)
    \end{pmatrix}.
\end{align}
For simplicity, we will consider that agents are equally rational ($\rational_A=\rational_B=\rational$) and that their moving rates are also identical ($\omega_A=\omega_B=\omega$).

We are looking for conditions to observe dynamical patterns and/or static phase separation. Notably, the homogeneous state is linearly unstable if one eigenvalue of $L$ has a positive real part. It is important to stress that the linear stability analysis is unable to predict the dynamic behavior when nonlinear terms become relevant. Whether the eigenvalues display an imaginary part or not \emph{does not} bring any information on the final dynamics of the system. For the sake of completeness, we explicitate the criteria to have eigenvalues with positive real part and zero imaginary part, referred to as case (i), and eigenvalues with positive real part and nonzero imaginary part, referred to as case (ii). We lie in case (i) if
\begin{align}
\begin{cases}
        \tr L > 0 \\
        (\tr L)^2 -4\det L>0,
\end{cases}
\text{ or }
\begin{cases}
    \tr L < 0\\
    \det L<0.
\end{cases}
\end{align}
Case (ii) is obtained if
\begin{align}
\begin{cases}
    \tr L>0 \\
    (\tr L)^2-4\det L<0.
    \end{cases}
\end{align}
The criterion $\tr L>0$ notably simplifies into
\begin{align}
    \bar \rho_A \partial_1 u_A +\bar \rho_B \partial_2 u_B> \frac{1}{\rational\hat G_\sigma (k)}\left( \frac{2-\rho_0}{1-\rho_0}  \right).
\end{align}
In the main text we have come up with utility functions that lead to eigenvalues with positive real parts and non zero imaginary parts, thus suggesting chasing instability. In some cases, oscillations were observed close to the homogeneous state but they eventually vanished at late times.  Whether or not the chasing instability or oscillations are sustained cannot be predicted from the simple linear stability analysis but would require to perform a weakly non-linear analysis which is beyond the scope of this present paper.

\section{LSA for two populations with local moves}
\label{app:LSA_two_populations_local}

We start from the local jump approximation of the mean-field equation for the coupled fields. We find that the dynamics can be cast into
\begin{equation}
    \partial_t \rho_A = \partial_x[\rho_A(1-\rho_A-\rho_B)\partial_x\mu([\rho_{A,B}],x)],
\end{equation}
with $\mu = \mu_{\mathrm{ent.}} + \mu_\mathrm{util.}$,
\begin{equation}
    \mu_\mathrm{ent.} = w_{\rational_A}(0) \log \left( \frac{\rho_A}{1 - \rho_A - \rho_B} \right),
\end{equation}
\begin{equation}
    \mu_\mathrm{util.} = -2 w_{\rational_A}'(0) u_A([\rho],x),
\end{equation}
and likewise for $\rho_B$. One can look into the stability of an homogeneous state with densities $\bar \rho_A$ and $\bar \rho_B$, expanding around this state with a utility $u(\phi_A, \phi_B)$ for agents $A$ and $v(\phi_A, \phi_B)$ for agents $B$. For convenience, we will take $ w_{\rational_A}(0)= w_{\rational_B}(0)=\omega f_\rational(0)=\omega/2$ and $w_{\rational_A}'(0)= w_{\rational_B}'(0)=\omega f_\rational'(0)=\omega\rational/4$.
Expanding around the homogeneous state $(\bar \rho_A$ , $\bar \rho_B)$ leads to
    \begin{equation}
        \begin{cases}
    \partial_t \rho_{A,1} = \frac{\omega}{2} \left[ (1-\bar \rho_B)\partial_x ^2 \rho_{A,1} + \bar \rho_A \partial_x ^2 \rho_{B,1} - \partial_x (\rational \bar \rho_A (1- \rho_0) \partial_x \phi_{A,1}\partial_1 u + \partial_x \phi_{B,1}\partial_2 u)  \right]\\
   \partial_t \rho_{B,1} = \frac{\omega}{2}\left[ (1-\bar \rho_A)\partial_x ^2 \rho_{B,1} +  \bar \rho_B \partial_x ^2 \rho_{A,1} - \partial_x (\rational \bar \rho_B (1- \rho_0) \partial_x \phi_{A,1}\partial_1 v + \partial_x \phi_{B,1} \partial_2 v) \right],
    \end{cases}
\end{equation}
Hence, in Fourier space, the linear system can be cast into
\begin{align}
    \partial_t \begin{pmatrix}
\hat \rho_A(k,t) \\
\hat \rho_B(k,t) 
\end{pmatrix}=K\begin{pmatrix}
\hat \rho_A(k,t) \\
\hat \rho_B(k,t) 
\end{pmatrix},
\end{align}
with 
\begin{align}
    K= \frac{\omega k^2}{2}
    \begin{pmatrix}
      \bar \rho_B-1 +\rational \bar\rho_A(1-\rho_0)\hat G_\sigma(k)\partial_1 u 
      & -\bar \rho_A +\rational \bar\rho_A(1-\rho_0)\hat G_\sigma(k)\partial_2 u \\
      -\bar \rho_B +(1-\rho_0)\bar \rho_B\rational \hat G_\sigma(k)\partial_1 v  
      & \bar \rho_A-1 +(1-\rho_0)\bar\rho_B\rational\hat G_\sigma(k) \partial_2 v
    \end{pmatrix}.
\end{align}
It is interesting to note that the evolution matrix $K$ is directly proportional to $L$ and, as a consequence, the stability criterion of the homogeneous state with local moves is exactly the same as the one found for non-local moves.
\end{widetext}

\section{Introducing a non-linear price field}
\label{app:complex_price_field}

\begin{figure}
\centering\includegraphics[width=\columnwidth]{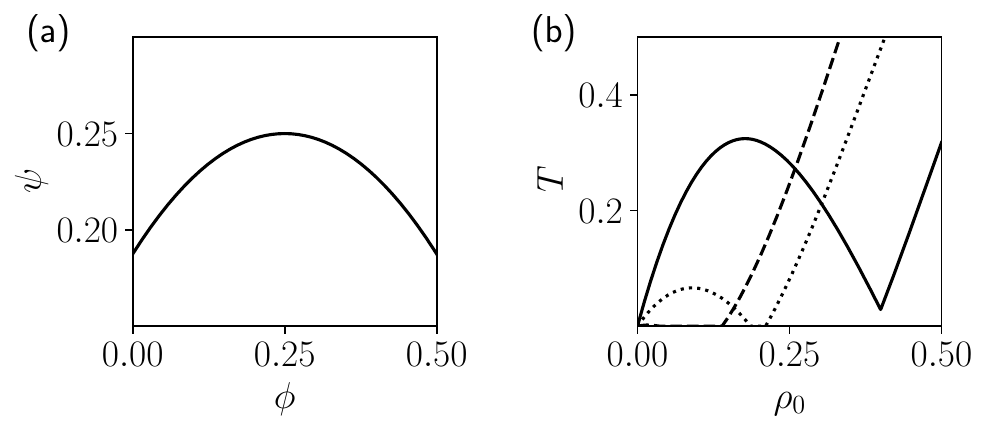}
    \caption{a) Non-monotonic price field $\psi(\phi)$ as a function of the locally perceived density $\phi$, given by Eq.~\eqref{eq:nonMonotonic_price_field}. Parameters: $\rho_\mathrm{p}^\star = 0.25$ and $\alphap = 2$. b) Spinodal curves for parameters $\lambda = 5, \alphap= 2, \alpha = 3$ and densities [$\rho^\star=0.2,\rho_\mathrm{p}^\star=0.4$] (solid line), [$\rho^\star=0.45,\rho_\mathrm{p}^\star=0.2$] (dotted line) and [$\rho^\star=0.5,\rho_\mathrm{p}^\star=0.1$] (dashed line). }
    \label{fig:price_field}
\end{figure}
For completeness, we can also consider a non-monotonic price field (see Fig.~\ref{fig:price_field}(a))
\begin{equation}
     \psi= \rho_\mathrm{p}^\star-\lvert \phi - \rho_\mathrm{p}^\star \rvert ^{\alphap}.
     \label{eq:nonMonotonic_price_field}
\end{equation} 
The intuition behind this relation is that prices can be lower in overcrowded neighborhoods as well as in empty neighborhoods, and are maximized for a given density $\rho_\mathrm{p}^\star$. Assuming again that the price-adjusted utility has the form $\bar{u}(\phi)= u(\phi) - u_\mathrm{p} [\psi(\phi)]$, with $u_\mathrm{p} [\psi]=\lambda\psi$ and $\lambda>0$, the total utility for the agents is then given by
\begin{equation}
    \bar{u}(\phi)= -\lvert \phi - \rho^\star \rvert ^\alpha + \lambda \lvert \phi - \rho_\mathrm{p}^\star \rvert ^\alphap +cst,
\end{equation} 
We can inject this expression into the linear-stability condition, see Eq.~\eqref{eq:LSA_general}, to pinpoint the condensation. In Fig.~\ref{fig:price_field}(b) we take $\alpha = 3$, $\lambda = 5, \alphap= 2$, and we interestingly observe that some spinodal curves display several re-entrance points.

\bibliography{schelling_AMB.bib}

\end{document}